\documentclass[twocolumn]{aastex6}
\pdfoutput=1

\usepackage{pdfpages}
\makeatletter
\AtBeginDocument{\let\LS@rot\@undefined}
\makeatother
\usepackage{amsmath}

\shorttitle{SFH of the starburst galaxies NGC~1705, NGC~4449 and
  Holmberg II} \shortauthors{Cignoni et al.}

\begin{document}


\title{Star Formation Histories of the LEGUS Dwarf Galaxies (I):\\
  recent History of NGC~1705, NGC~4449 and Holmberg
  II\footnote{Based on
  observations obtained with the NASA/ESA \textit{Hubble Space
    Telescope} at the Space Telescope Science Institute, which is
  operated by the Association of Universities for Research in
  Astronomy under NASA Contract NAS 5-26555.}}

\author{M. Cignoni$^{1,2,3}$, E. Sacchi$^{3,4}$, A. Aloisi$^{5}$,
  M. Tosi$^{3}$, D. Calzetti$^{6}$, J. C. Lee$^{5,7}$, E. Sabbi$^{5}$,
  A. Adamo$^{8}$, D. O. Cook$^{19}$, D. A. Dale$^{9}$,
  B. G. Elmegreen$^{10}$, J.S. Gallagher III$^{18}$,
  D. A. Gouliermis$^{11,12}$, K. Grasha$^{6}$, E. K. Grebel$^{13}$,
  D. A. Hunter$^{14}$, K. E. Johnson$^{17}$, M. Messa$^{8}$,
  L. J. Smith$^{15}$, D. A. Thilker$^{16}$, L. Ubeda$^{5}$ and
  B.C. Whitmore$^{5}$}

\affil{
$^{1}$Dipartimento di Fisica, Universit\`a di Pisa, Largo Bruno Pontecorvo, 3, 56127 Pisa, Italy\\
$^{2}$INFN, Sezione di Pisa, Largo Pontecorvo 3, 56127 Pisa, Italy\\
$^{3}$INAF--Osservatorio di Astrofisica e Scienza dello Spazio, Via Gobetti 93/3, I-40129 Bologna, Italy\\
$^{4}$Dipartimento di Fisica e Astronomia, Universit\`a degli Studi di Bologna, Via Gobetti 93/2, I-40129 Bologna, Italy\\
$^{5}$Space Telescope Science Institute, 3700 San Martin Drive, Baltimore, MD 21218, USA\\
$^{6}$Department of Astronomy, University of Massachusetts -- Amherst, Amherst, MA 01003, USA\\
$^{7}$Visiting Astronomer, Spitzer Science Center, Caltech, Pasadena, CA, USA\\
$^{8}$Department of Astronomy, The Oskar Klein Centre, Stockholm University, Stockholm, Sweden\\
$^{9}$Department of Physics and Astronomy, University of Wyoming, Laramie, WY\\
$^{10}$IBM Research Division, T.J. Watson Research Center, Yorktown Hts., NY\\
$^{11}$Zentrum f\"ur Astronomie der Universit\"at Heidelberg, Institut f\"ur Theoretische Astrophysik, Albert-Ueberle-Str.\,2, 69120 Heidelberg, Germany\\
$^{12}$Max Planck Institute for Astronomy,  K\"{o}nigstuhl\,17, 69117 Heidelberg, Germany\\
$^{13}$Astronomisches Rechen-Institut, Zentrum f\"ur Astronomie der Universit\"at Heidelberg, M\"onchhofstr.\ 12-–14, 69120 Heidelberg, Germany\\
$^{14}$Lowell Observatory, Flagstaff, AZ\\
$^{15}$European Space Agency/Space Telescope Science Institute, Baltimore, MD\\
$^{16}$Department of Physics and Astronomy, The Johns Hopkins University, Baltimore, MD\\
$^{17}$Department of Astronomy, University of Virginia, Charlottesville, VA \\
$^{18}$ Dept. of Astronomy, University of Wisconsin--Madison, Madison,
WI\\
$^{19}$ California Institute of Technology, Pasadena, CA\\
}

\begin{abstract}
 
  We use HST observations from the Legacy Extragalactic UV Survey to
  reconstruct the recent star formation histories (SFHs) of three
  actively star-forming dwarf galaxies, NGC~4449, Holmberg II and
  NGC~1705, from their UV color-magnitude diagrams (CMDs). We apply a
  CMD fitting technique using two independent sets of stellar
  isochrones, PARSEC-COLIBRI and MIST, to assess the uncertainties
  related to stellar evolution modelling. Irrespective of the adopted
  stellar models, all the three dwarfs are found to have had almost
  constant star formation rates (SFRs) in the last 100-200 Myr, with
  modest enhancements (a factor of $\sim$2) above the 100
  Myr-averaged-SFR. Significant differences among the three dwarfs are
  found in the overall SFR, the timing of the most recent peak and the
  SFR$/$area. The Initial Mass Function (IMF) of NGC~1705 and Holmberg
  II is consistent with a Salpeter slope down to $\approx$ 5
  M$_{\odot}$, whereas it is slightly flatter, s$=-2.0$, in NGC~4449.

  The SFHs derived with the two different sets of stellar models are
  consistent with each other, except for some quantitative details,
  attributable to their input assumptions. They also share the
  drawback that all synthetic diagrams predict a clear separation in
  color between upper main sequence and helium burning stars, which is
  not apparent in the data. Since differential reddening, significant
  in NGC~4449, or unresolved binaries don't appear to be sufficient to
  fill the gap, we suggest this calls for a revision of both sets of
  stellar evolutionary tracks.

\end{abstract}



\section{Introduction}

Dwarf galaxies are important. Given the ubiquity of low metallicity
galaxies at high redshift, understanding how star formation (SF)
proceeds in such regimes is crucial for cosmological purposes. Dwarf
galaxies lack spiral density waves and differential rotation, hence,
global dynamics presumably plays a minor role in regulating their
SF. Moreover, whilst massive galaxies have formed most of their stars
in the first $\sim 3$ Gyr, low mass galaxies have been forming stars
over the whole of cosmic time. This results in a variety of star
formation histories (SFH, see e.g. \citealt{THT09,weisz11,
  gallart15,mq11}) and specific star formation rates (sSFR), ranging
from nearly inactive, as the transition-type dwarf Leo~T, to extremely
active, as the blue compact dwarfs (BCDs).

  Potential mechanisms for triggering SF in dwarfs can be broadly
  categorized as either internally-driven or externally-driven events.
  From the internal point of view, intrinsic processes such as
  stochastic self-propagating SF (\citealt{gerola80}) have certainly a
  role, in particular at the low mass end where single star forming
  events can change the galaxy luminosity significantly. However, BCDs
  appear to be different from non-starbursting dwarf irregulars in
  terms of their HI and internal dynamics. The higher central
  rotation-velocity gradients of BCDs suggests a relation between the
  intense activity and the gravitational potential
  (\citealt{lelli12}). Moreover, despite the isolation, BCDs show star
  forming regions kinematically decoupled from the rest of the galaxy
  (\citealt{koleva14}). \cite{elme12} suggested that giant star
  forming clumps in dwarf irregular galaxies could in-spiral towards
  the center, feeding intense star formation. This is because in such
  small galaxies, star forming clumps could be sufficiently massive to
  exceed a few percent of the galaxy mass enclosed inside their
  orbital radii, therefore producing dynamically significant torques
  on dark matter halo particles, halo stars, and the surrounding disk
  to lose their angular momentum with a timescale of 1 Gyr. Other
  proposed internal mechanisms are the effect of massive triaxial dark
  haloes, promoting the migration of a significant fraction of gas
  from the periphery to the center of the dwarf (\citealt{bekki02}),
  or bars made of dark matter (\citealt{he04}).


External processes must be also important. If Cold Dark Matter (CDM)
is scale-free as predicted, groups of dwarf galaxies are a natural
expectation on small mass scales and dwarf galaxies could be formed
through the same physical process as massive galaxies. While dwarfs
accreting dwarfs have been observed in at least one case, DDO~668
(\citealt{annibali16,sacchi16}), dwarf groups have been recently
observed by \cite{stier17} using the panchromatic TiNy Titans (TNT;
\citealt{stier15}) survey, a systematic study of SF in interacting
dwarf galaxies. According to this study, the interaction between dwarf
galaxies could be quantitatively different as compared to more massive
counterparts. In fact, both paired dwarfs and paired massive galaxies
show enhanced SF out to separations of $\sim 100$ kpc
(e.g. \citealt{patton13,stier15}), but the effect in dwarfs is
stronger by a factor of 1.3 and involves a larger fraction of the
virial radius. Dwarfs are therefore more globally affected by small
companions. A large fraction of interacting dwarfs are starbursting,
too. If one defines as starbursts those regions with global
H${\alpha}$ equivalent width $> 100\,$ \r{A}, \cite{stier17} found
that starbursts occur in 20\% of the TNT isolated pairs, compared to
only 6\% of the matched isolated single dwarfs, suggesting that
dwarf-dwarf interactions play a key role in triggering starbursts in
dwarfs. Interestingly, the final coalescence stage of the dwarf-dwarf
merger is not required to produce starbursts, but rather they are
triggered at earlier stages of the interaction. Finally, collisions
between dwarfs and satellites predominantly star-free, or even dark,
could explain the strong bursts of SF that in some dwarfs occur
without apparent cause (\citealt{stark16}). { Indeed, HI clouds
  with no evident optical counterpart have been found around several
  BCDs (\citealt{ramya09}), suggesting that infall of such gas clouds
  is not an unlikely event.}


{ In this framework, } quantitative information on the SFH in dwarf
galaxies (see e.g. \citealt{weisz11}) is fundamental in order to
understand the connection between the observed activity and processes
such as merging, accretion, and interaction. The SFH likely keeps a
record of the past interactions. Detailed high resolution SFH studies
may also provide invaluable hints on how star-forming regions evolve
and what turns off SF, from the smallest to galactic-size scales, and
how these processes depend on dynamical effects.

The purpose of this paper is to quantify the most recent SFH of
individual galaxies using UV Color-Magnitude Diagrams (CMDs) of their
resolved stars.  While all studies of SFH performed so far
(e.g. \citealt{THT09} and references therein, \citealt{weisz11}) aim
at covering the longest possible look-back time and infer the SFH back
to the earliest epochs, in this case we aim at resolving in higher
detail the most recent epochs.  Here we exploit the Legacy
ExtraGalactic UV Survey (LEGUS; \citealt{calzetti15}) to study at high
temporal resolution the most recent past of three actively
star-forming dwarf galaxies, NGC~1705, NGC~4449 and Holmberg II
(hereafter Ho~II, also known as UGC~4305).  { All three dwarfs have
  their optical SFH already derived by others (see e.g.,
  \citealt{annibali03,weisz08,mq10}), but our study is complementary
  to theirs since it deals only with the last 100-200 Myr but with the
  higher temporal resolution achieved with UV photometry.  Moreover,
  we derive the SFHs adopting two different and independent sets of
  isochrones, to estimate the uncertainties resulting from different
  approaches and input physics (rotation, mixing length, amount of
  overshooting, etc.) of the stellar models. We also take advantage of
  the higher capability of separating main-sequence (MS) and post-MS
  stars in UV CMDs to evaluate how well the evolutionary models are
  able to reproduce the observed properties of massive stars, an
  important test for stellar theories.}

\begin{table*}
\centering
\caption{Fundamental properties of NGC~4449, Ho~II and NGC~1705.}
\label{infogen}
\begin{tabular}{lcccccccc}
\hline
\hline
Galaxy&RA       &DEC     &$D$  &E(B-V)&$M_{HI}$          &$M_{\star}$     &$12+\log(O/H)$& M$_{B}$\\
           &(J2000)&(J2000)&(Mpc)&   &$(10^7$M$_{\odot})$&
                                                           $(10^7$M$_{\odot})$& & \\
\hline
 \object{NGC~4449}&$12\,44\,42.7$&$-35\,58\,00$&$3.82\pm0.18^{a}$&$0.20^{a}$& $300\pm77^{g}$& $110^{i}$&$8.26^{d}$ & $-18.2^{l}$\\
 \object{Ho~II}&$12\,46\,00.4$&$-33\,50\,17$&$3.38^{b}$&$0.03^{b}$ & $73^{h}$ &$23^{i}$ &$7.92^{f}$&$-16.7^{m}$ \\
 \object{NGC~1705}&$12\,54\,53.6$&$-28\,20\,27$&$5.1\pm0.6^{c}$& $0.045^{c}$&$11.1\pm2.9^{g}$ & $13^{i}$&$7.96-8.28^{e}$&$-15.45^{n}$ \\
 \hline
\end{tabular}\\
(a)\cite{annibali08};
(b)\cite{dalcanton12};
(c)\cite{tosi01};
(d)\cite{berg12};
(e)\cite{mou10};
(f)\cite{crox09};
(g)\cite{lelli14};
(h)\cite{hr89};
(i)\cite{calzetti15} and references therein;
(l)\cite{hunter99};
(m)\cite{tully88};
(n)\cite{annibali15}.
\end{table*}

Table \ref{infogen} describes the relevant properties of these
galaxies that could be gathered from the literature.  { The three
  galaxies show different degrees of isolation, with NGC~1705 being
  the most isolated and NGC~4449 the least isolated.}  The most
massive one, NGC~4449, is similar in size and mass to the Large
Magellanic Cloud (see e.g. \citealt{kar13}). This galaxy has been
classified as a ``global'' starburst, since SF appears to involve much
of the galaxy (\citealt{hunter97}). Its closest neighbour is at a
projected distance of 41 kpc and previous studies
(\citealt{mdg12,rich12}) have reported evidences of interactions and
of a stellar tidal stream. The second most massive system, Ho~II, is
similar to the Small Magellanic Cloud in terms of absolute magnitude,
HI and total mass (\citealt{puche92}). VLA HI observations
(\citealt{puche92}) have shown a large number of HI shells and holes
in the ISM, probably caused by stellar feedback from multiple
generations of SF spread out over tens or hundreds of Myr
(\citealt{weisz09}). The closest neighbour of Ho~II, Kar52, is at a
projected distance of 30 kpc. The least massive dwarf, NGC~1705, is
also the most gas-poor and isolated (its closest neighbours, LSBG
F157-089 and MRSS 157-121650, are at more than 500 kpc; see
e.g. \citealt{firth06}, \citealt{evsti07}). This galaxy is the only
one hosting a super star cluster (SSC; M$_{B} = -14.5$) and exhibiting
strong galactic winds (see e.g. \citealt{meurer92}). The isolation
makes NGC~1705 an ideal candidate to understand how long the SF can be
sustained in isolation and whether the activity is linked to the
accretion of cold gas from the inter galactic medium. Indeed, cold
flows of gas are predicted by $\Lambda$CDM models of galaxy formation
(\citealt{keres15}). NGC~1705 is also the only one morphologically
classified as a BCD, while the other two are ``irregulars''.

The SFHs are recovered using the synthetic CMD approach applied to the
F336W vs F336W-F555W CMD. This filter combination is ideal for
identifying the youngest populations in our galaxies, mostly traced by
main sequence (MS), Hertzsprung gap (HG) and core He-burning (HeB) stars
more massive than 5 M$_{\odot}$. While we could have used the bluest
LEGUS filter available, the F275W, improving the age resolution in the
last 50 Myr, this would have exacerbated the effect of reddening
canceling the resolution gain. Moreover, it would have shortened the
look-back time reachable by the photometry.

This paper is organised as follows. In Sect. 2 we describe
observations and data reduction. In Sect. 3 we discuss the spatial
distribution and we compare the observed CMDs with stellar evolution
models. In Sect. 4 we describe the SFH recovery technique. In Sect. 5
we present the recovered SFHs. Finally, Sect. 6 and 7 summarize our
results.

\section{Observations}

The observations are part of the LEGUS survey, an {\it{HST}}
panchromatic stellar survey of 50 nearby star forming dwarf and spiral
galaxies. The goal of the survey is to investigate scales and modes of
SF using the leverage of the UV imaging.  Scientific objectives and
the data reduction are described in \cite{calzetti15}, while stellar the
photometry is described in detail in { Sabbi et al. 2018}. The
observations were performed with the Wide Field Camera 3 (WFC3) and
Advanced Camera for Surveys (ACS) in parallel, in a set of broad bands
over the range 0.27 - 0.81 $\mu$m, respectively F275W, F336W, F438W,
F555W and F814W (equivalent to NUV, U, B, V, and I respectively) and
combined with archival optical ACS data.

Resolved stellar photometry was performed using the DOLPHOT package
(e.g., Dolphin 2000). DOLPHOT performs point-spread function (PSF)
fitting on all the flat-fielded and CTE-corrected images (FLC) per
field, simultaneously. We then used the main DOLPHOT routine to make
photometric measurements in each filter independently on the
pre-processed images, relative to the coordinate system of the
drizzled reference image. The output photometry from DOLPHOT is on the
calibrated VEGAMAG scale (see \citealt{sabbi18}).



\section{Data}

\subsection{Spatial distributions}

\begin{figure*}[t!]
\centering \includegraphics[width=18cm]{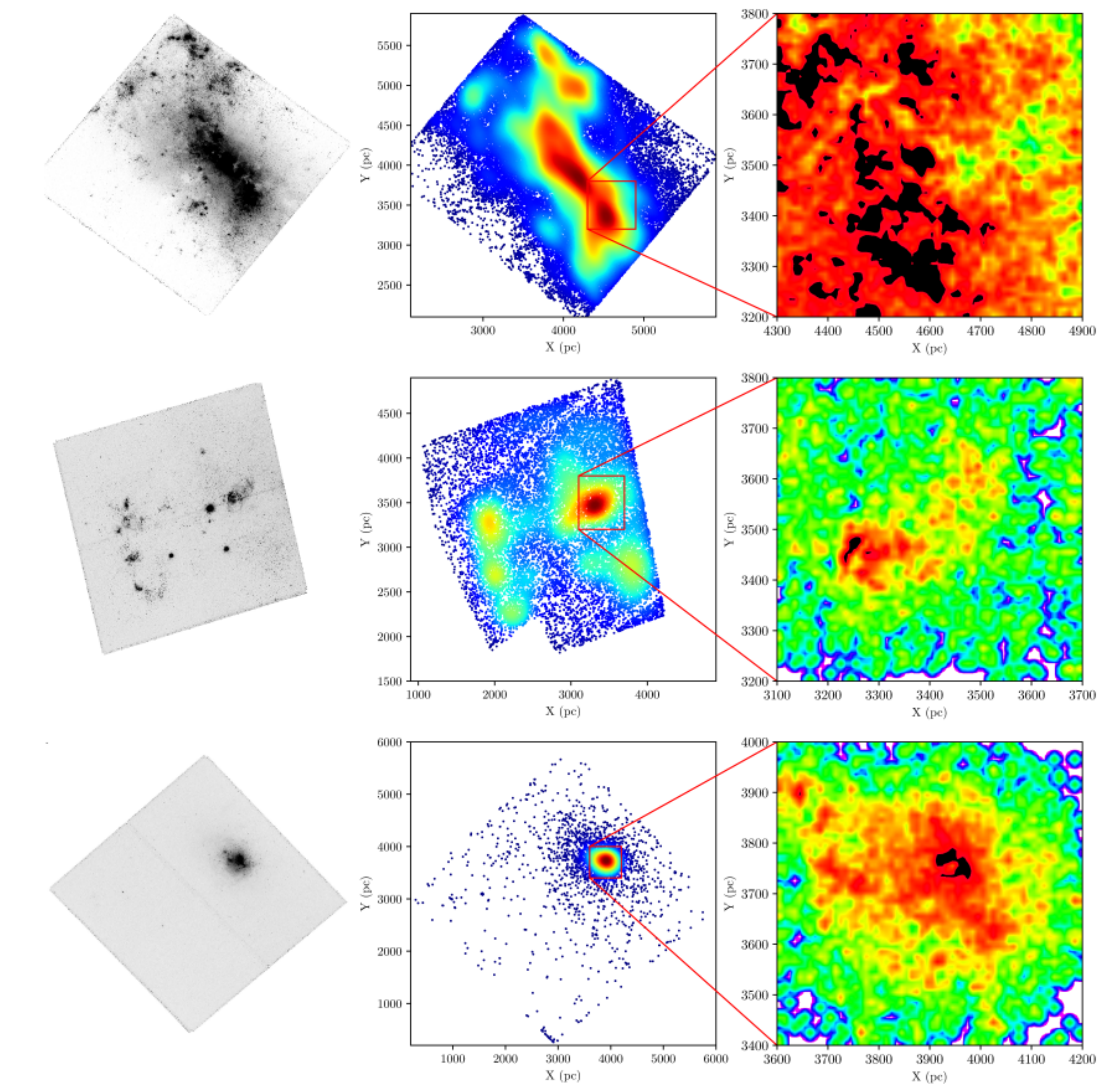}
\caption{Left column: F336W images of NGC~4449 (top panel),
  Ho~II (middle panel) and NGC~1705 (bottom panel). Middle column: all
  stellar sources detected in F336W and F555W, color-coded with
  the density (estimated with a Gaussian kernel). Right column:
  zoom-in on the densest sub-regions.}
\label{map} 
\end{figure*}

Figure \ref{map} shows the spatial distributions of the resolved stars in the three
galaxies (from top to bottom: NGC~4449, Ho~II and NGC~1705): the
left column shows the F336W images, the middle column shows the map in
physical units (pc) of all stellar sources identified both in the
F336W and in the F555W, color coded with the density estimated using a
Gaussian kernel ({ standard deviation 0.2 pc}), and the right column
shows zooms-in of the densest sub-regions ($\approx\, 600$
pc wide). In order to make the plots easily comparable, color maps are
normalized to the same limits.

The three galaxies have very different morphologies. NGC~1705
is characterised by a single roundish overdensity, whereas NGC~4449
and Ho~II have multiple overdensities that are irregularly
distributed. Zooming in on the densest sub-regions (right panels)
reveals other differences. NGC~4449's peak sub-region is very crowded
(as indicated by the almost uniform red color), with multiple peaks
(black areas) of higher density. Compared to NGC~4449, Ho~II's
subregion has a much lower average density (most of the area is
green), with only three peaks at the same density of
NGC~4449. NGC~1705's subregion has intermediate properties between
NGC~4449 and Ho~II, except in the very center where it is even denser
than NGC~4449 (the very center is extremely crowded, as witnessed by
the ``ring'' shape of the black region\footnote{This is certainly
  related also to the presence of the SSC at the center of NGC~1705.},
a symptom that we are missing stars in the center). In conclusion, on
average, NGC~4449, NGC~1705 and Ho~II represent a sequence of
decreasing average density. On the other hand, when only the peak
density is considered, the sequence from the highest to the lowest
becomes NGC~1705, NGC~4449 and Ho~II.

\subsection{CMDs}

\begin{figure*}[t]
\centering \includegraphics[width=18cm]{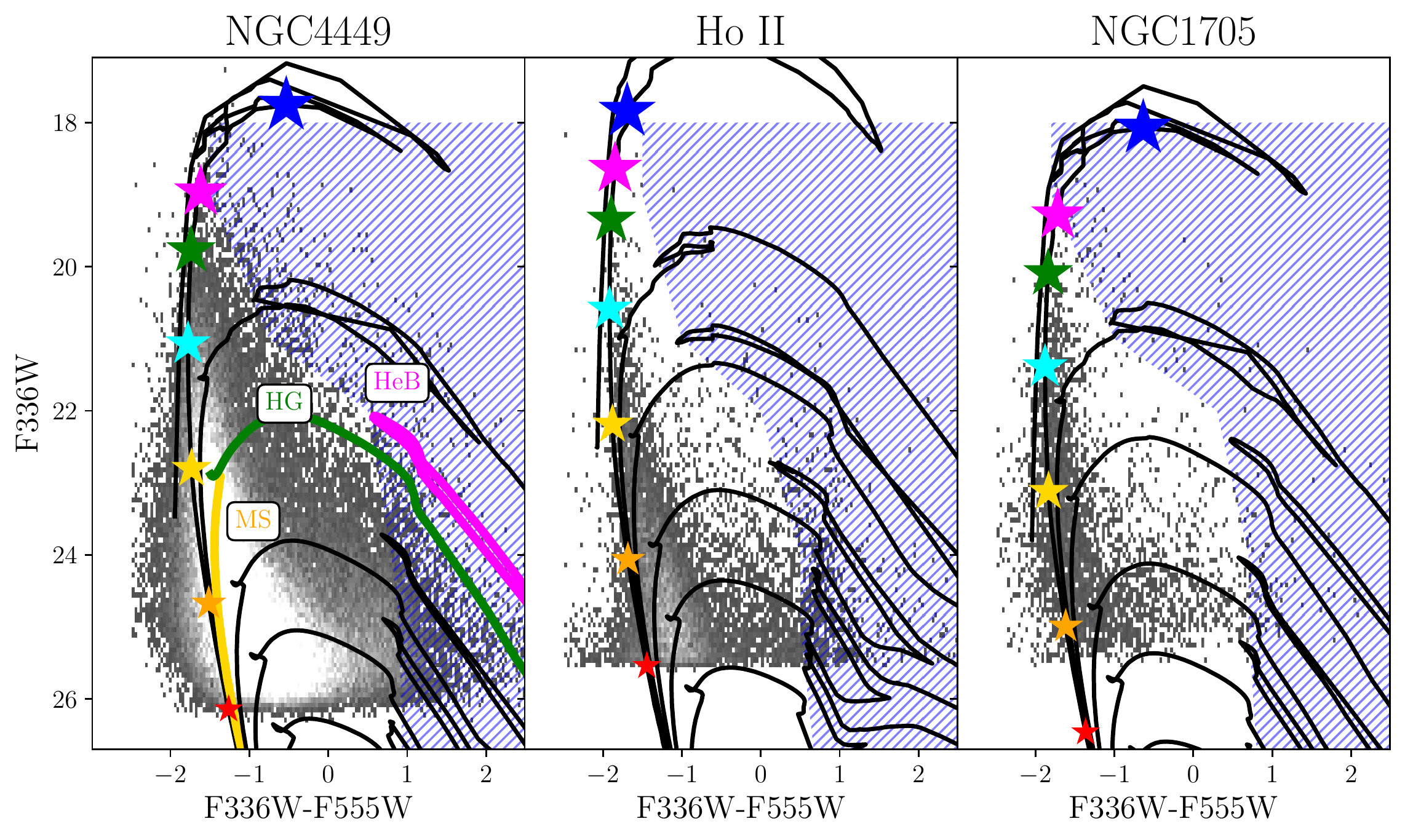}
\caption{From left to right: CMDs for NGC~4449, Ho~II and
  NGC~1705. PARSEC-COLIBRI isochrones of ages 3, 10, 20, 50, 100, 200
  Myr are overlaid. The most relevant stellar phases are also
  indicated along the 20 Myr isochrone (leftmost panel) with lines of
  different colors: orange for MS, green for HG, and magenta for
  HeB. { The adopted distance modulus, foreground reddening
    E(B$-$V) and metallicity Z of each galaxy are respectively: 27.9,
    0.1, 0.005 (for NGC~4449), 27.6, 0.03, 0.002 (for Ho~II), 28.5,
    0.045 and 0.005 (for NGC~1705).  The hatched area indicates the
    region of the CMD populated by HeB stars. Star symbols represent
    PARSEC-COLIBRI stellar models of 3 Myr and masses 6 (red), 10
    (orange), 20 (gold), 40 (cyan), 60 (green), 70 (magenta), 80
    (blue) M$_{\odot}$.}}
\label{3cmd_iso} 
\end{figure*}

In order to reject non-stellar objects and to have a clean final
sample of stars for our CMDs, we applied quality cuts. The DOLPHOT
output was filtered to only allow objects classified as stars with
signal-to-noise $> 5$ in both filters. { In addition, we selected
  detections for which DOLPHOT reports an error flag of 3 or less. The
  list was further culled using sharpness ($< 0.15$) and crowding
  ($< 1.3$).  Our final star catalogs for NGC~4449, Ho~II and NGC~1705
  contain 107761, 11516 and 3741} stars, respectively.  The resulting
``cleaned'' CMDs are shown in Fig. \ref{3cmd_iso} together with the
latest PARSEC-COLIBRI isochrones (PAdova and TRieste Stellar Evolution
Code version 1.2S plus COLIBRI code for AGB thermal pulses;
\citealt{bressan12,tang14,marigo17}) for the ages 3, 10, 20, 50, 100
and 200 Myr { (corrected for reddening E(B$-$V) and distance moduli
  0.1 and 27.9, for NGC~4449, 0.03 and 27.6, for Ho~II, 0.045 and
  28.5, for NGC~1705).} The adopted metallicity is Z$=$ 0.005 for
NGC~4449 and NGC~1705, and 0.002 for Ho~II.

Guided by the isochrones, we recognise three common stellar phases in
the CMDs of Fig. \ref{3cmd_iso}: a prominent blue plume of MS stars, a
HG and a HeB phase. Typical masses populating these CMDs are larger
than 5 M$_{\odot}$ and as massive as 80 M$_{\odot}$. Given the depth
of these CMDs the maximum age that can be investigated in NGC~4449,
Ho~II and NGC~1705 is about 180 Myr, 200 Myr and 100 Myr,
respectively.

Besides the similarities, these CMDs also show some marked differences
in morphology that cannot be fully explained by observational
effects. Most notably, the MS of NGC~4449 appears broader than the one
of either NGC~1705 or Ho~II. Moreover, the luminosity distribution of
HG/HeB stars appears: 1) rather uniform in NGC~4449; 2) declining
with increasing luminosity in Ho~II; 3) clumpy in NGC~1705.

Concerning the MS spread, it seems unlikely that a metallicity
variation can explain it all, despite the higher metallicity of
NGC~4449. In fact, in massive hot stars the opacity is dominated by
electron scattering, which depends only on the mass fraction of
hydrogen $X$ as $0.20\,(1 + X)$, and is about constant at Z $<$
0.01. Moreover, the observed MS in Ho~II and NGC~1705 is bracketed by
the 3 Myr isochrone, which encompasses the blue edge, and the MS
termination point (MSTP; the massive analogue of the MS Turn-Off) of
all isochrones, defined by the red extent of MS evolution (visible as
a hook in all isochrones older than 3 Myr). This feature suggests that
age is the dominant driver of the MS width in Ho~II and NGC~1705, rather
than photometric errors (modest), unresolved binaries or differential
reddening.

On the other hand, the MS of NGC~4449 is significantly
broader\footnote{The MS width is measured with the standard deviation
  of the color distribution of stars in the magnitude interval
  $22<$F336W$<23$.} than expected from an age spread alone. Given
the youth of the population, it appears very probable that
differential reddening is artificially inflating the apparent width of
the MS. Indeed, using far-ultraviolet and Balmer-line imagery for 22
OB complexes in NGC~4449, \cite{hill94} derived reddening values E(B$-$V) ranging from 0.25 to
0.6 from the Balmer
decrement. Alternatively, stellar rotation (see, e.g., \citealt{mm97}) or larger
core-convective overshooting (see, e.g., \citealt{bressan15}) could be
also responsible for a broadening the MS. However, such effects should occur also in HoII and NGC~1705, whereas only in NGC~4449 the MS
is broader than expected from an age spread.

{ For intermediate/massive stars beyond the MS}, the stellar phase
before the red giant branch (RGB) is so fast ({ Kelvin-Helmholtz
  timescale}) that the probability of observing stars is low compared
to nuclear phases. Generally, this causes the so-called Hertzsprung
gap, i.e. the observed lack (or paucity) of stars in the evolutionary
phase right after the MS. However, the stellar production in these
galaxies is such that the number of stars in these fast evolutionary
phases is not negligible. Indeed, HG stars are copious in NGC~4449,
with no sign of interruption between F336W$\approx 19$ and the faint
end of the CMD. In contrast, the bulk of HG stars in Ho~II is fainter
than F336W$\approx 21$, while the distribution of HG stars in NGC~1705
seems bimodal, with overdensities around F336W$\approx 21$ (part of
these objects could be HeB stars) and below F336W$\approx 22$.

Concerning the HeB phase, theory predicts that the bluest point in the
blueward excursion of the core HeB phase for masses above the 2.5
M$_\odot$ (the so called ``Blue Loop'', BL) depends on several
quantities, including stellar mass (more massive stars having more
extended loops), metallicity (more metal-rich stars having less
extended blue loops) and input physics (see e.g. \citealt{tang16}). On
the other hand, the luminosity of the BL is mainly driven by the
stellar mass. In fact, the energy budget during the HeB phase is
provided by the HeB core and the H-burning shell. Since above 2.5
M$_\odot$ the He-core increases with increasing total mass as a
consequence of the increasing mass of the convective core during the
previous core H-burning phase, the HeB luminosity depends on the mass
of the star, providing a useful mass-luminosity relation (and, in
turn, a luminosity-age relation).

The comparison with the isochrones in Fig. \ref{3cmd_iso} suggests
that in all three galaxies the blue edge of the BL is redder than
F336W$-$F555W$\approx 0.5$ { for stars fainter than 22 (see the
  hatched area) and reaches F336W$-$F555W$\approx -1.5$ at magnitudes
  around 19}. Clearly, the relatively high metallicity of these
galaxies (Z$=0.002-0.005$) prevents the HeB phase to reach the MS
color. The net result is that, depending on the magnitude, stars at
intermediate colors are a different mixture of HG and BL stars. In
Ho~II and NGC~1705, the CMD region located between $18<$F336W$< 21$
and $-0.5 <$F336W$-$F555W$<$0 is likely populated by both HG and BL
stars, whereas the CMD region between $21<$F336W$< 25$ and
F336W$-$F555W$\approx 0$ is mostly populated by HG stars. The
situation is more ambiguous in NGC~4449, where the color spread tends
to smear out many CMD features already at magnitudes as bright as 24.

Since HG and BL monotonically fade as the population age increases,
both stellar species are fundamental clocks to infer the recent SFH of
a galaxy. The route of the BL as age indicator has been pioneered by
\cite{dp} on the dwarf irregular galaxy Sextans A and recently applied
to several other dwarf irregulars by, e.g., \cite{mq11,mq12}.

The main advantage of HG/BL over the MS is that subsequent generations
of HG/BL stars \emph{do not overlap each other} as they do on the
MS. Moreover, the HG/BL is on average \emph{brighter} than the coeval
MSTP. However, since the MS evolutionary times are much longer than
post-MS times, starcounts along the MS are statistically more
robust. This property becomes crucial in galaxies like NGC~1705 and
Ho~II, whose low global SFR does not provide a sufficient number of
post-MS stars. In the next Sections we will use the HG and BL stars in
synergy with MS stars to infer the recent SFH of the three dwarfs.

\begin{figure*}[t]
\centering \includegraphics[width=13cm]{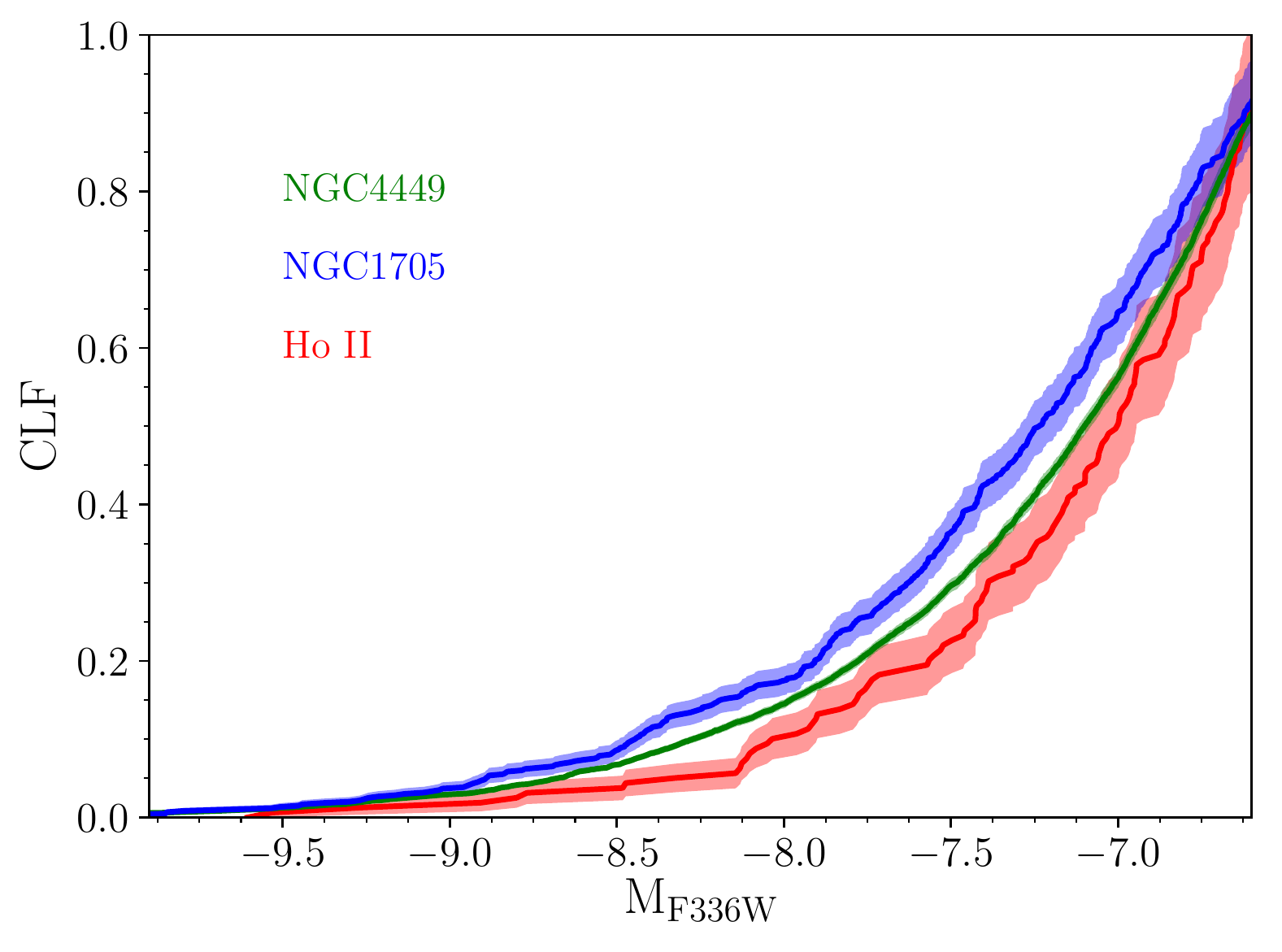}
\caption{Cumulative Luminosity Functions, in absolute magnitude, for NGC~4449 (green solid
  line), Ho~II (red solid line) and NGC~1705 (blue solid line). The
  shaded envelopes represent the Poisson uncertainty ($1\,\sigma$).}
\label{cumul_LF} 
\end{figure*}

Before closing this Section we demonstrate that the comparison of
the luminosity function (LF) for the three dwarfs provides a measure
of the relative SFRs. To do this, since the total number of stars in
NGC~4449 is significantly higher than in the other two, we rely on the
cumulative LF (CLF).  Fig. \ref{cumul_LF} shows the CLFs for the three
galaxies corrected for the corresponding distance modulus and
reddening E(B-V) (see Table \ref{infogen}), { normalized to the
  number of stars with absolute magnitude M$_{\mathrm{F336W}}$
  brighter than $-6.5$\footnote{ This limit corresponds to an
    approximate apparent magnitude F336W$= 21.5$, where starcounts are
    complete in all three galaxies.  } (corresponding to stars younger
  than $\approx \, 15 $ Myr)}. The CLF of NGC~1705 increases faster
than the other two, which are similar to each other (compatible within
$1\,\sigma$ as shown by the error bars). Since the number of stars
brighter than a certain magnitude is proportional to the average SFR
younger than a certain age, the CLF is also a measure of the relative
SFR (normalized to the rate 15 Myr ago). Under this light,
Fig. \ref{cumul_LF} suggests that NGC~1705 is relatively younger than  Ho~II and
NGC~4449, despite the latter being globally more
massive. Indeed, above M$_{\mathrm{F336W}}=-6.5$ NGC~4449 hosts five
times more stars than NGC~1705, and thirty times more stars than
Ho~II.  We suggest that the recent activity in these
galaxies is not driven by the total mass but rather by external
factors or stochastic phenomena.

\section{Synthetic color-magnitude diagrams}

\subsection{Artificial star tests}

A mandatory ingredient for meaningful data-model comparisons is to
quantify photometric errors and incompleteness of the data. To this
aim, a typical approach is a proper artificial star test, i.e. the
injection of ``fake'' sources (one at a time) onto the actual images:
the source detection routine used for our science images is applied to
the fields containing the combined science images and the fake
sources. Counting how many fake stars are lost as a function of
magnitude and position provides the map of the local incompleteness.

Since young stars ($<\,100-200$) Myr tend to be concentrated in
associations and filaments, most of them will be affected by more
significant errors and incompleteness than the average in the
field. To overcome this problem we followed the procedure described by
\cite{cigno16}. As a first step, fake stars are evenly distributed
over the galaxy field of view (FoV). The observed density of stellar sources (plagued
by incompleteness) is then corrected for the local incompleteness,
restoring the ``true'' profile of the galaxy. As a second step, fake stars are injected following
the reconstructed profile. The resulting incompleteness, being
weighted with the real stars, will be an unbiased estimate of the
actual incompleteness suffered by the young stars.

\begin{figure*}[t]
\centering \includegraphics[width=19cm]{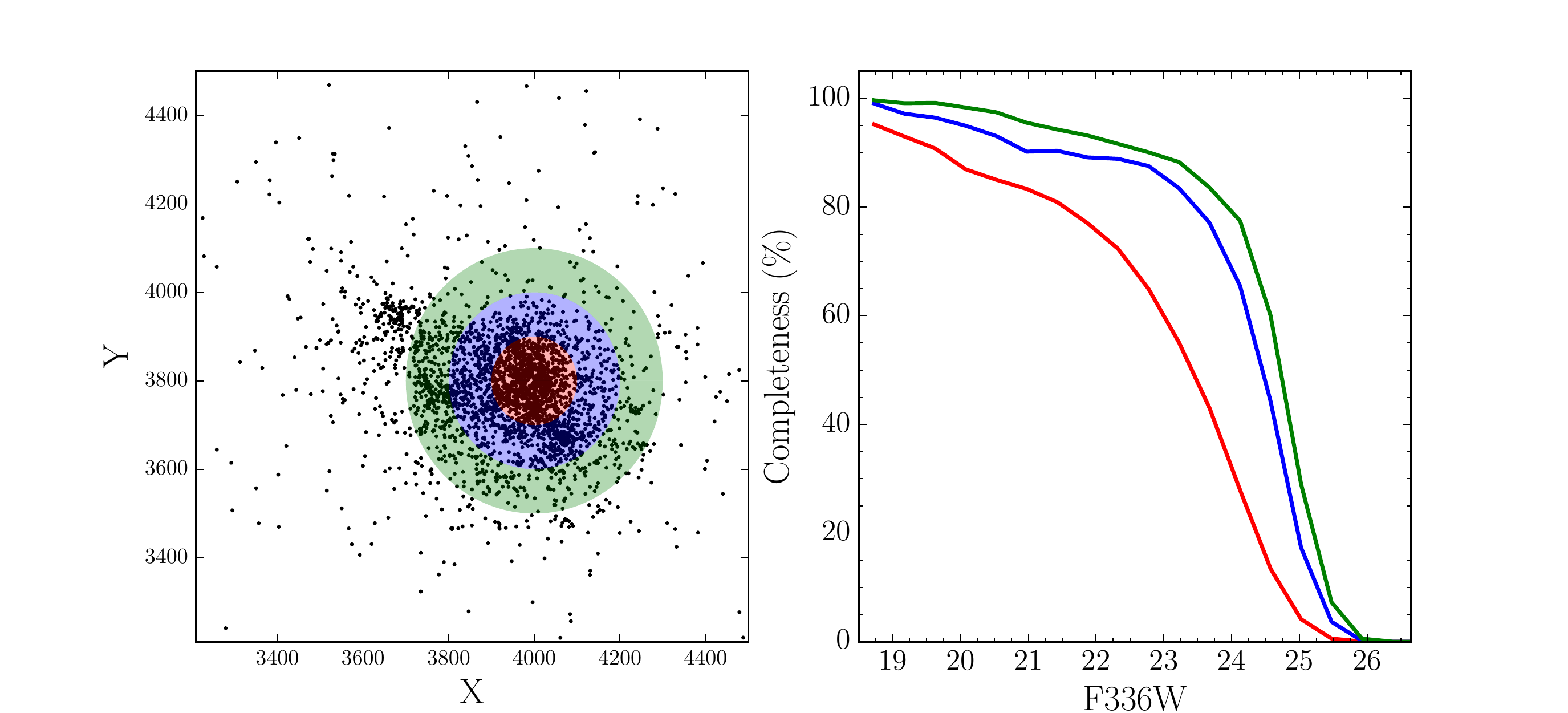}
\caption{Photometric completeness (right panel) in the F336W in three
  concentric annuli (left panel) around the center of NGC~1705. The red
  one refers to the stars within 100 pixel (1 pixel is about 1 pc at
  the distance of NGC~1705) from the center, the blue one between 100
  and 200 pixel (about), the green one between 200 and 300 pixel.}
\label{compl_1705} 
\end{figure*}

As an example, Fig. \ref{compl_1705} shows the estimated F336W
completeness profiles (right panel) of NC1705 in three concentric
regions (see left panel) centred in the densest region of the
galaxy. As expected, where crowding is most severe completeness drops
faster. In the central region (red line) the F336W catalogue is 50\% incomplete
at 23.5, while in more external annuli a 50\% incompleteness is
reached at 24.5 (blue line) and 24.75 (green line).

\subsection{SFH recovery algorithm}

The SFH of the three galaxies was recovered using the population
synthesis code SFERA (Star Formation Evolution Recovery
Algorithm). The analysis details are described in
\cite{cigno15,cigno16}. Here, we provide only a short description of
the procedure; a complete discussion is available in the quoted papers
above.

A library of ``basic'' synthetic CMD is generated using the isochrones
from the adopted sets of stellar models. To test the systematic
uncertainties due to the stellar models, in our procedure the
synthetic CMDs are generated adopting either the PARSEC-COLIBRI or the
MIST (MESA Isochrones and Stellar Tracks ) isochrones
(\citealt{paxton11,paxton13,paxton15,choi16,dotter16}).  The two sets
differ both in the input physics and in the assumptions about the
efficiency of macroscopic mechanisms, like rotation (PARSEC-COLIBRI
models are static, MIST are rotating with $v/v_{crit}=0.4)$. Each
basic synthetic CMD is a Monte Carlo realization with constant SFH (in
a given bin of ages) and fixed metallicity ($\pm\, 0.05$ dex). In this
work we adopted 9 logarithmic time bins, namely
$\log({\mathrm{age}}) = 5.0-6.6,\, 6.6-7.0,\, 7.0-7.2,\, 7.2-7.4,\,
7.4-7.6,\, 7.6-7.8,\, 7.8-8.0,\, 8.00-8.25,\, 8.25-8.50$. Concerning
metallicity, we allowed the code to use metallicities\footnote{To be consistent with their own assumptions, we adopt the approximation [M/H]$=\log(Z/Z_{\odot})$, with
  $Z_{\odot}=0.0152$ for PARSEC-COLIBRI models and $Z_{\odot}=0.0142$
  for MIST models.}  between $[M/H] = -2$ and 0.1 with a resolution of
0.1 dex. Once the metallicity and ages are selected, each basic CMD is
populated with a \cite{kroupa01} initial mass function (IMF) between
0.1 and 300 M$_{\odot}$. { Other IMF slopes are considered if no
  SFH is found to match the data. The last step concerns the binary
  population: 30\% of synthetic stars are considered to be part of
  unresolved binary systems and their flux is combined with a
  companion, whose mass is a random fraction between 0 and 1 of the
  primary.}

Concerning distance and extinction we adopted the following approaches.
Given the relatively low sensitivity of the UV SFH to the assumed
distance (mainly due to the rapid evolutionary phases populating the
UV CMD), we opted to use literature distances (see Table
\ref{infogen}) based on the tip of the RGB (see
e.g. \citealt{bellazzini01}), a robust feature missing in the UV CMDs
but clearly visible in optical CMDs. On the other hand, the high
sensitivity of the UV color to the reddening distribution allowed us
to model reddening and SFH simultaneously. The extinction distribution
is modeled with two parameters: a foreground extinction applied evenly
to all stars, AV, and a differential extinction, dAV\footnote{Each
  star is attributed a random additional extinction between 0 and dAV.},
that is applied in addition to AV. Moreover, we adopted the extinction
law of \cite{cardelli89} and assumed a normal total-to- selective
extinction value of Rv$=3.1$.

To mimic the observational effects and uncertainties, each synthetic CMD is
then convolved with photometric errors (derived from the cumulative
distribution of mag$_{\mathrm{out}}$-mag$_{\mathrm{input}}$ of fake
stars) and incompleteness as derived from the artificial star tests.

\begin{figure*}[t]
\centering \includegraphics[width=16cm]{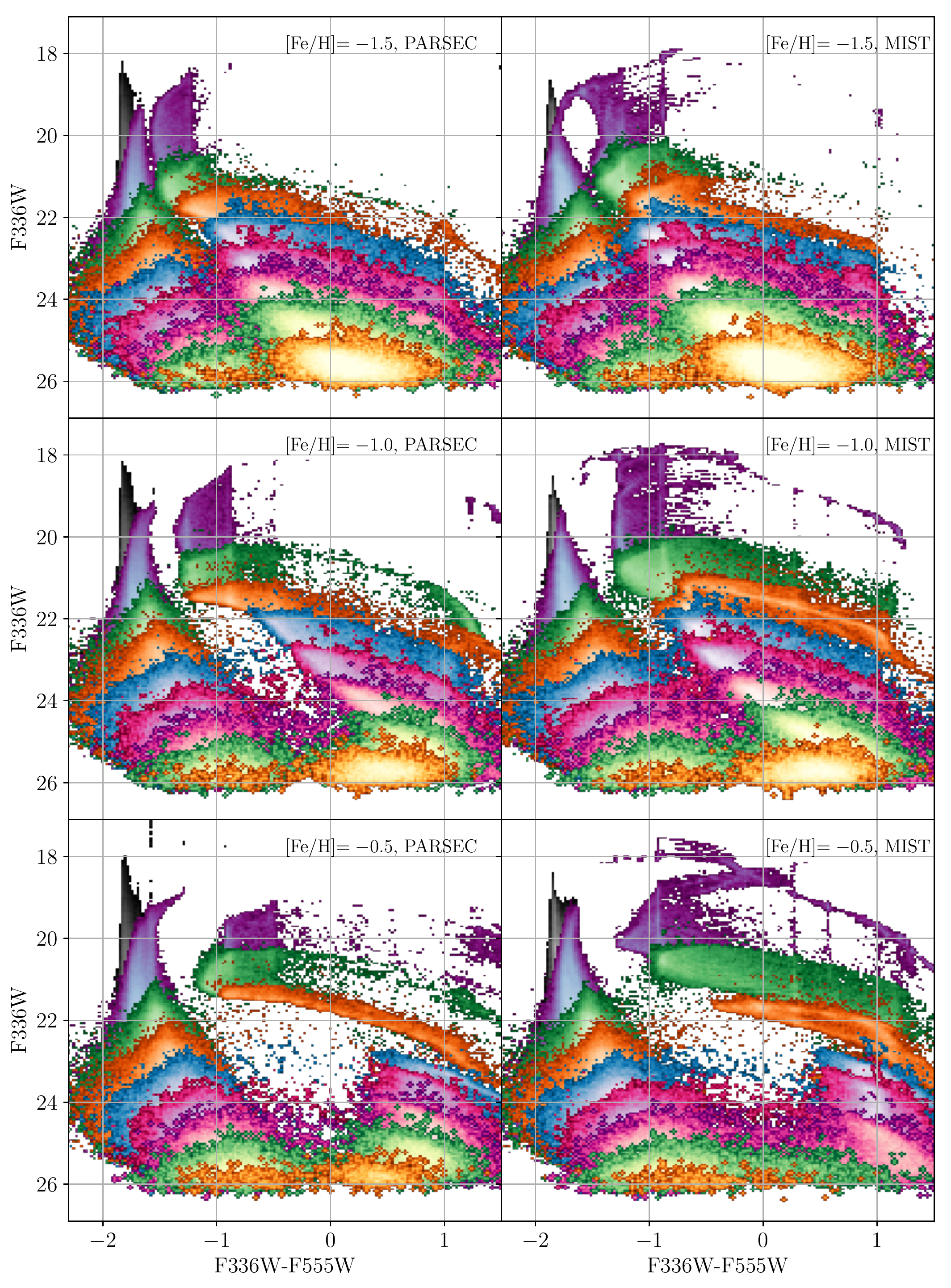}
\caption{From top to bottom, example of basic synthetic CMDs generated
  for $[\mathrm{M/H}]=-1.5,-1.0,-0.5$, errors/incompleteness of
  NGC~4449, binary fraction of 30\% and distance modulus of NGC~4449
  (see Table 1). { For each metallicity, different colors represent
    different star formation episodes with logarithmic duration
    $\log(\mathrm{age}) = 5.0-6.6,\, 6.6-7.0,\, 7.0-7.2,\, 7.2-7.4,\,
    7.4-7.6,\, 7.6-7.8,\, 7.8-8.0,\, 8.00-8.25,\, 8.25-8.50$. The
    adopted distance and foreground extinction are 27.9 and
    E(B$-$V)=0.1, respectively.} PARSEC-COLIBRI models are on the
  left, MIST models are on the right.  }
\label{6cmd} 
\end{figure*}

Finally, the combination of basic CMDs that minimizes the residuals
with respect to the observational CMD (in terms of Poissonian
likelihood) is searched with the hybrid-genetic algorithm
SFERA. Figure \ref{6cmd} shows an example of basic synthetic CMDs
(PARSEC-COLIBRI on the left, MIST on the right) for three different
metallicities ($[\mathrm{M/H}]=-0.5,-1.0,-1.5$).
At fixed metallicity, the differences between PARSEC-COLIBRI and MIST
synthetic CMDs are indicative of the different underlying stellar
physics and stellar evolution codes. When deriving the SFH in the next Section, these differences
will serve as an indication of the potential differences between a
given model and the ``truth''.

Figure \ref{6cmd} also confirms the low dependence upon metallicity in
this range of ages. The only significant effect concerns the color
extension of the BL, which shrinks with increasing metallicity. Given
this difficulty of establishing the metallicity, we opted to use the
available spectroscopic information (see Table 1) together with the
photometry to provide additional constraints on the SFH. In
particular: 1) the metallicity in the last 10 Myr was fixed at the
spectroscopic value of the observed HII regions (within the measured errors); 2) we limited the
number of free parameters by imposing that the metallicity was an
increasing function of time.

Throughout this work, we adopt the oxygen abundance as a tracer of the
overall recent metallicity (in the last 10 Myr).

\section{SFHs}

\subsection{Ho~II}

\begin{figure*}[t]
\centering \includegraphics[width=13cm]{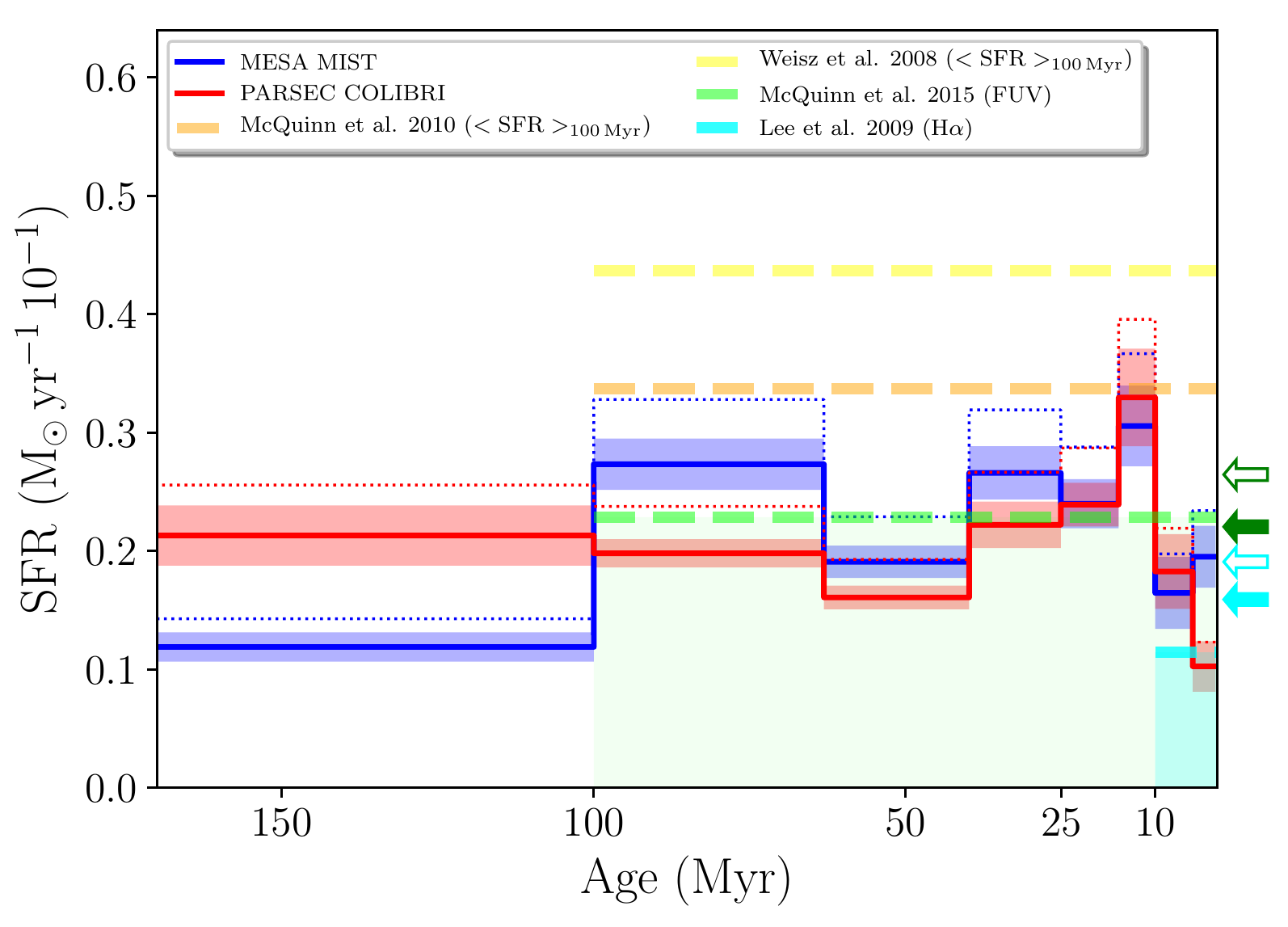}
\caption{Recovered SFH of Ho~II. Blue and red continuous lines
  represent the MIST and PARSEC-COLIBRI solutions, respectively. {
    The adopted IMF is Kroupa-2001. Thin dotted lines are
    the corresponding solutions obtained extrapolating the IMF
    exponent $-2.3$ down to 0.1
    M$_{\odot}$. The green and cyan filled arrows on the right side of
    the plot stand for our 100 Myr-averaged-SFR and 10
    Myr-averaged-SFR (averaged between the two solutions),
    respectively (open arrows are extrapolation using an exponent
    $-2.3$ down to 0.1 M$_{\odot}$). The orange and yellow dashed
    lines show the 100 Myr-averaged-SFR found by, respectively, Weisz
    et al. 2008 and McQuinn et al. 2010. The green dashed line is the
    average rate computed using FUV emission by \cite{mq15}, while the
    cyan dashed line indicates the average rate from H$\alpha$
    observations computed using the SFR(FUV)$/$SFR(H$\alpha$)
    conversion factor computed by \cite{lee09}.}}
\label{sfr_ugc4305} 
\end{figure*}

Figure \ref{sfr_ugc4305} shows the recovered SFHs for Ho~II. { Red
  and blue continuous lines are for PARSEC-COLIBRI and MIST models,
  respectively, while the dotted lines are the same SFHs extrapolated
  with IMF exponent s$=-2.3$ down to 0.1 M$_{\odot}$. The green and
  cyan filled arrows on the right side of the plot line indicate our
  100 Myr-averaged-SFR and 10 Myr-averaged-SFR (averaged between the
  two solutions), respectively (open arrows are extrapolation using an
  exponent $-2.3$ down to 0.1 M$_{\odot}$). Orange, yellow, green and
  cyan thick dashed lines are instead SFRs taken from the literature
  (see below).}  Overall, results obtained using the two sets of
isochrones are qualitatively similar. This is not surprising because the CMD of Ho~II is dominated by MS stars, for which different
evolutionary models generally agree. Both solutions are consistent
with an almost constant SF activity over the last 180 Myr, with mild
enhancements whose intensity is at most 1.5 higher than the 100
Myr-averaged-SFR { ($0.022\,$M$_{\odot}/$yr; see the green filled
  arrow on the right side of the plot)}. The SFH obtained with MIST
models is less constant, with a broad enhancement between 60 and
100 Myr ago. The SFH with PARSEC-COLIBRI models shows a similar recent
peak, whereas the general activity is more constant. The best
foreground extinction AV is $0.25$ mag, while the differential one dAV
is $0.02$ mag.

\cite{dalcanton12} found similar recent SF enhancements using HST/ACS
optical CMDs. They derived the relative SFH (normalized to the
lifetime averaged rate) for two different regions dubbed UGC4305-1 and
UGC4305-2. The latter is the one with larger overlap with our UV
observations, so we used its SFH for comparison with our solution. The
main difference between the two is the timing of the most recent peak,
which is younger than 10 Myr in their solution whereas it is between
10 and 16 Myr ago in ours. The ratio between the peak rate and the
average activity in the last 200 Myr is instead very similar, with
Dalcanton's et al. 2012 ratio only slightly higher (2 vs 1.5).

\cite{weisz08} derived the optical SFH for the entire area covered by
the HST/ACS observations (UGC4305-1 plus UGC4305-2) and found that
Ho~II has been about two times more active in the last 20 Myr than in
the last 200 Myr. { A similar behavior is also found by \cite{mq10}
  using the same data and approach. The yellow and orange thick dashed
  lines in Fig. \ref{sfr_ugc4305} show their 100 Myr-averaged-SFRs
  reduced by a factor 2.25\footnote{This number was estimated using
    the optical CMDs (Sabbi et al. 2018) for the two HST/ACS
    pointings.}, which is the ratio between the number of bright stars
  (F555W$<24$) found in the two ACS pointing and in our WFC3 pointing
  (we also note that both \citealt{weisz08} and \citealt{mq10} used a
  Salpeter IMF down to 0.1 M$_{\odot}$). After this correction, these
  rates are only slightly higher than our average rate extrapolated
  with an IMF exponent $-2.3$ down to 0.1 M$_{\odot}$ (open arrows on
  the right side of Fig. \ref{sfr_ugc4305}). Moreover, the effect of
  using different stellar models (our models are based on the most
  updated evolutionary tracks) and filters (our F336W filter combined
  with the F555W provides a better { temperature} resolution for
  blue stars compared to F555W$-$F814W) could explain the slightly
  older age for our most recent peak.

  Another useful comparison is with current SFR indicators like
  H$\alpha$ nebular emission and far ultraviolet (FUV) non-ionizing
  continuum emission. While the first indicator arises from the
  recombination of gas ionized by the most massive O- and early-type
  B-stars, therefore tracing the SF in the last few million years, the
  UV flux stems from the photospheres of O- through later-type
  B-stars, and thus traces the SF in the last 100 Myr.  In
  Fig. \ref{sfr_ugc4305} the average SFRs derived from FUV and
  H$\alpha$ emission are shown with green and cyan thick dashed lines,
  respectively (both rates assume a Salpeter IMF down to 0.1
  M$_{\odot}$). The former combines the SFR$-$FUV calibration from
  \cite{k98} and the GALEX FUV flux, cropped to the footprint of the
  HST/ACS observations (taken from \citealt{mq15}) and corrected for
  the factor 2.25. The latter uses the ratio SFR(H$\alpha$)$/$SFR(FUV)
  provided by \cite{lee09}. Overall, both the SFR(FUV) and
  SFR(H$\alpha$) are in excellent agreement with our findings.}

\begin{figure*}[t]
\centering \includegraphics[width=15cm]{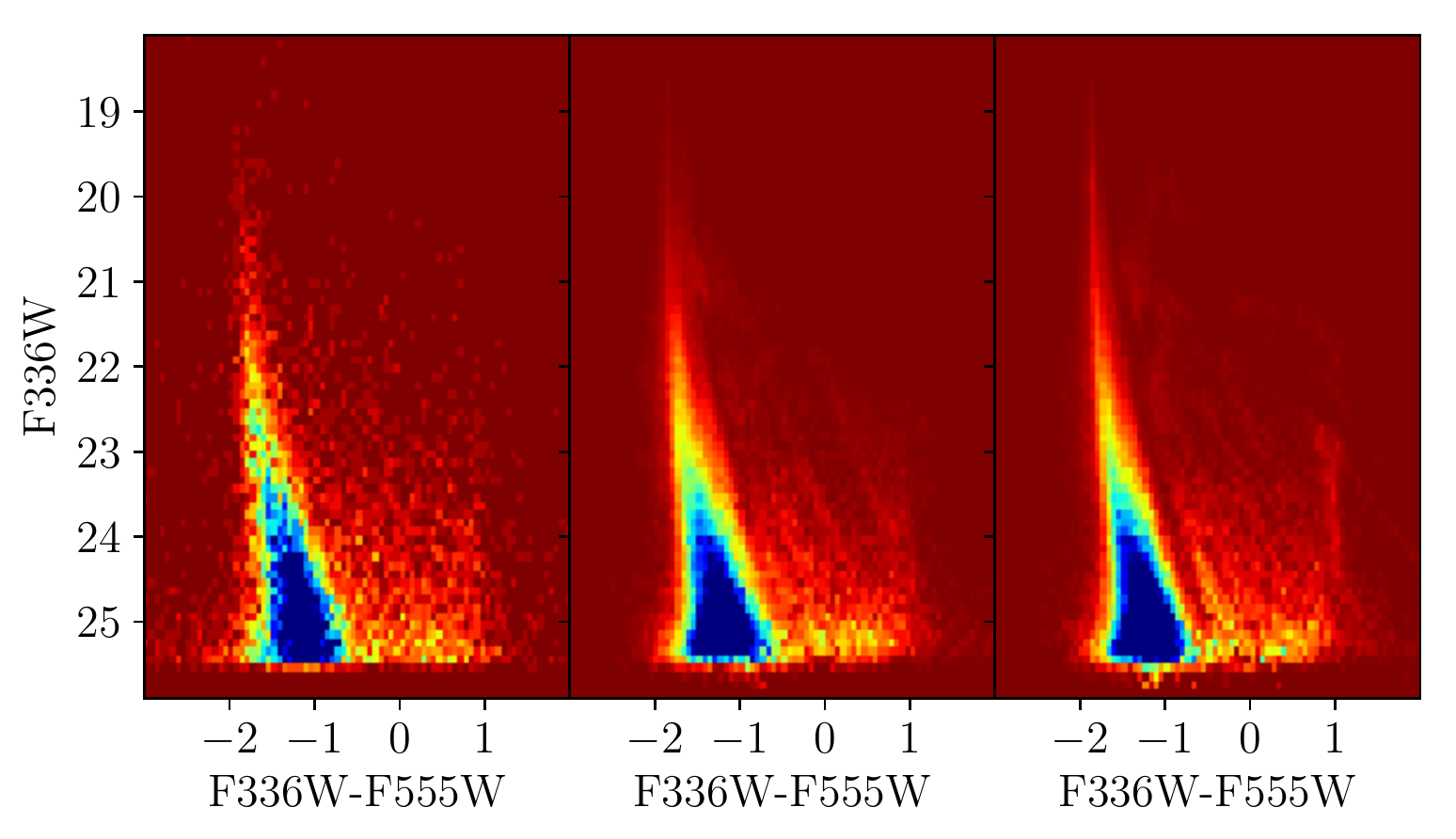}
\caption{Ho~II's CMD (left panel) compared to the best synthetic CMDs
  generated with PARSEC-COLIBRI (middle panel) and MIST (right panel)
  isochrones. }
\label{3cmd_ugc4305} 
\end{figure*}

\begin{figure}[t]
\centering \includegraphics[width=9cm]{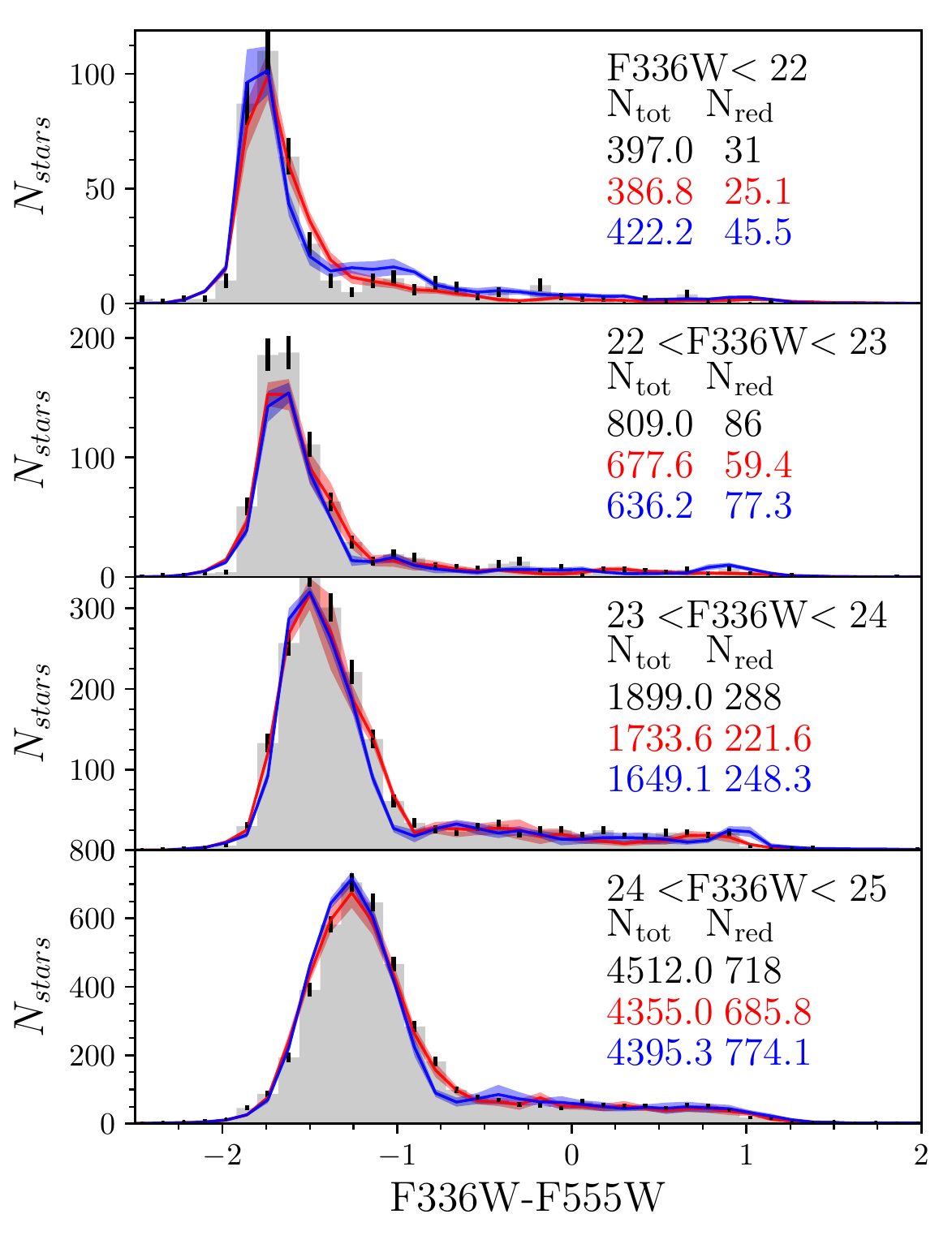}
\caption{Color distribution of Ho~II (grey shaded area), best PARSEC-COLIBRI
  model (red line) and best MIST model (blue line) for the
  labelled magnitude ranges. Observed and predicted counts (total and
  redder than color -0.5) are also shown.}
\label{CF_ugc4305} 
\end{figure}

The most basic check on the reliability of the recovered SFH is to
compare synthetic and observed CMDs. Figure \ref{3cmd_ugc4305} shows
the best synthetic CMDs (the middle panel corresponds to the
PARSEC-COLIBRI simulation, while the right-hand panel corresponds to
the MIST one) corresponding to the recovered SFH compared to the data
(left hand panel).

The color distribution of data vs simulations in four bins of
magnitude is shown in Fig. \ref{CF_ugc4305}.  Both models
reproduce very well the observations, with only a few appreciable
misfits. The colors and in particular the width of the simulated MS
match fairly well those of the observed one, in all the magnitude
bins.  The total number and color distribution of stars redder than
F336W$-$F555W$=-0.5$, likely bona fide { HG and BL stars}, are also
well matched. { For this galaxy, a power law IMF with exponent
  s$=-2.3$ down to 5 M$_{\odot}$ is consistent with the data.}


For the post-MS, we notice some mild issues in the colour
distribution. In the brightest bin of Fig. \ref{CF_ugc4305} (top
panel), the model distribution shows an excess of HeB stars with
colors between $-1$ and $-1.4$ (corresponding to the clump of stars at
F336W $=20-21$ in the synthetic CMDs of
Fig. \ref{3cmd_ugc4305}). Another general difference concerns how the
post-MS phase is populated. In the CMDs of Fig. \ref{3cmd_ugc4305} it
is clear that model stars tend to lie either near or far from the MS, whereas the observed post-MS distribution is
rather continuous in color.

It is unclear what triggered the increases in SF, since Ho~II is a
rather isolated galaxy, at the very edge of the M81 group, almost
detached from it. There are no signs of interaction between Ho~II and
close-by members of the M81 group, Kar 52 and UGC 4483 are located at
a projected distance of $\sim 30$ kpc and $\sim 105$ kpc,
respectively. From an orbital point of view, taking into account the
relative radial velocity between Ho~II and Kar 52, an interaction
might have happened 600 Myr ago, so differential rotation should have
dissipated the effects of a weak interaction.  { Moreover, since
  the dynamical mass of Ho~II is a factor of 100 greater than that of
  Kar 52 (\citealt{mi94}), the latter is unlikely to have affected
  significantly Ho~II. }

Using FUV, H$\alpha$, and HI observations, \cite{stewart00} studied
the possible internal mechanisms triggering SF activity in Ho~II. The authors
found that the energy deposited into the ISM from supernovae and
stellar winds are sufficient to account for the observed properties of
the HI distribution.

From deep VLA data, \cite{BC02} found that the distribution of neutral
hydrogen has a comet-like morphology – compressed on one side with a
faint extended component on the opposite side - with the tail pointing
away from the centre of the M81 group. The authors suggested that ram
pressure from a hot intragroup medium could be responsible for the HI
morphology, although tidal stripping could not be ruled out.

More recently, using deep, wide-field optical Subaru images
\cite{bernard12} studied the stellar populations in the outskirts of
Ho~II. They found very few, if any, Ho~II stars beyond a
galactocentric distance of $\sim 7$ arcmin, where the HI is instead
present and becomes distorted. Since tidal forces would affect gas and
stars equally, they argued that ram pressure is the likely culprit of
the HI morphology. Under this light, the interaction between the ISM
and the inter-galactic medium (IGM) could be responsible for the
recent SFR enhancement. Indeed, some ram pressure models (see
e.g. \citealt{bekki03}) indicate that the pressure of the IGM can
induce the collapse of molecular clouds and consequently trigger a
burst of SF. \cite{bernard12} also found old RGB stars more centrally
concentrated than young blue stars, which follow the (compressed) HI
contours.

\subsection{NGC~4449}

\begin{figure*}[t]
\centering \includegraphics[width=13cm]{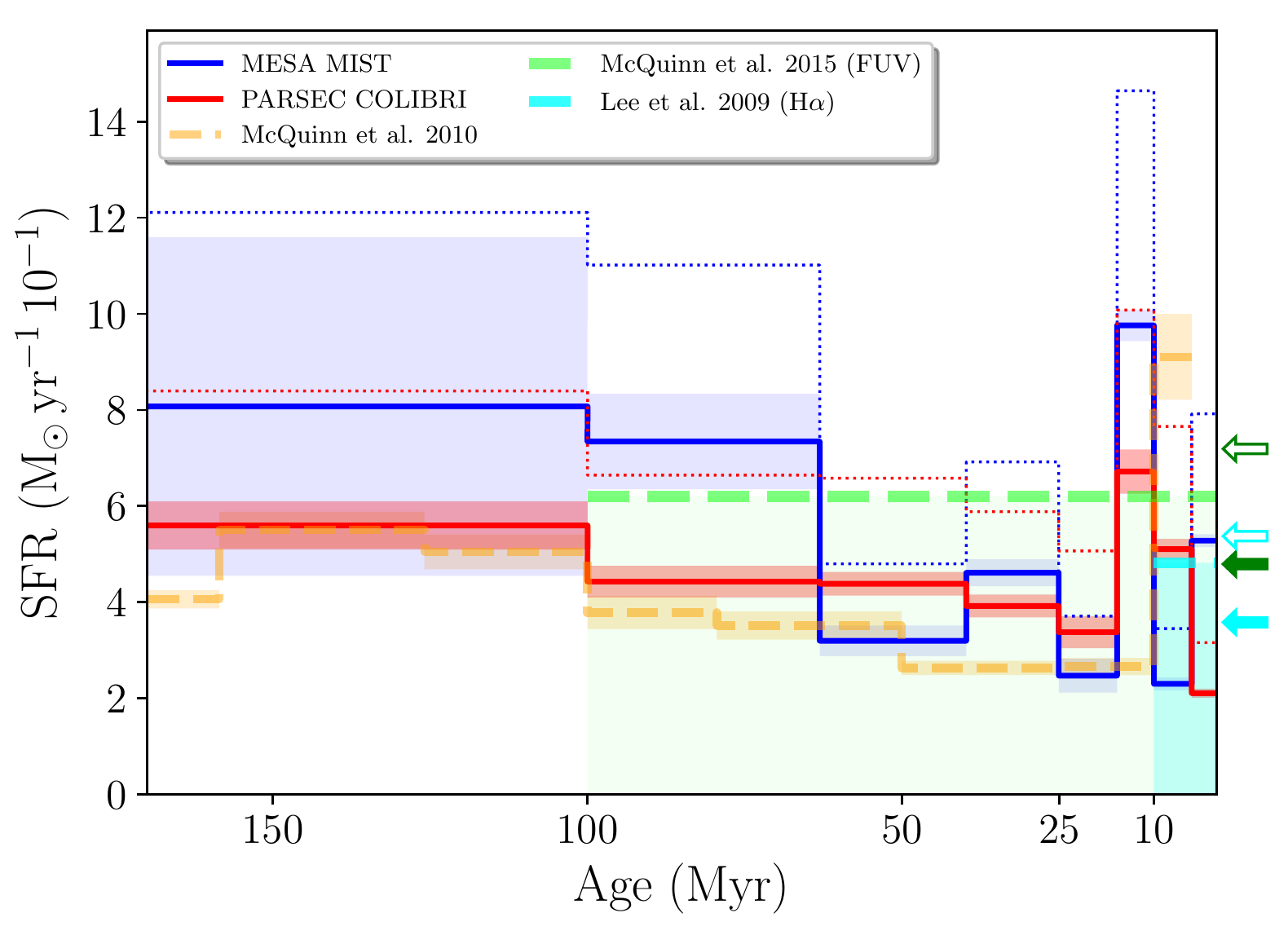}
\caption{Recovered SFH of NGC~4449. The adopted IMF is a double power
  law with exponent s$=-2.0$ above 0.5 M${_\odot}$, and s$=-1.3$
  below. Symbols are the same as in Fig. \ref{sfr_ugc4305}. }
\label{sfr_ngc4449} 
\end{figure*}

The SFHs in better agreement with the observational CMD are shown in
Figure \ref{sfr_ngc4449} ({ red and blue continuous lines are for
  PARSEC-COLIBRI and MIST models, respectively}). This galaxy
experienced a significant event between 10 and 16 Myr ago, when its
activity increased by a factor of $\sim 1.5-2$ above the 100
Myr-averaged-SFR { (indicated with a filled green arrow on the
  right side of the plot). SF has continued since that peak event at a
  lower level, consistent with the activity prior the peak. The best
  extinction value AV is $0.18$ mag, while the differential one dAV is
  $1.09$ mag. This solution was obtained using an IMF exponent
  s$=-2.0$ (instead of the canonical s$=-2.3$) above 5 M$_{\odot}$,
  since no combination of SFH and extinction matching the number of
  stars brighter than F336W$=23$ was found with s$=-2.3$: depending on
  the adopted stellar model, the latter underpredicts counts for
  magnitudes brighter than F336W$=23$ by 15 \% (MIST) and 23 \%
  (PARSEC-COLIBRI ). Given the lack of constraints below 5
  M$_{\odot}$, we arbitrarily chose to use s$=-2.0$ down to 0.5
  M$_{\odot}$, and s$=-1.3$ below.}


\begin{figure*}[t]
\centering \includegraphics[width=13cm]{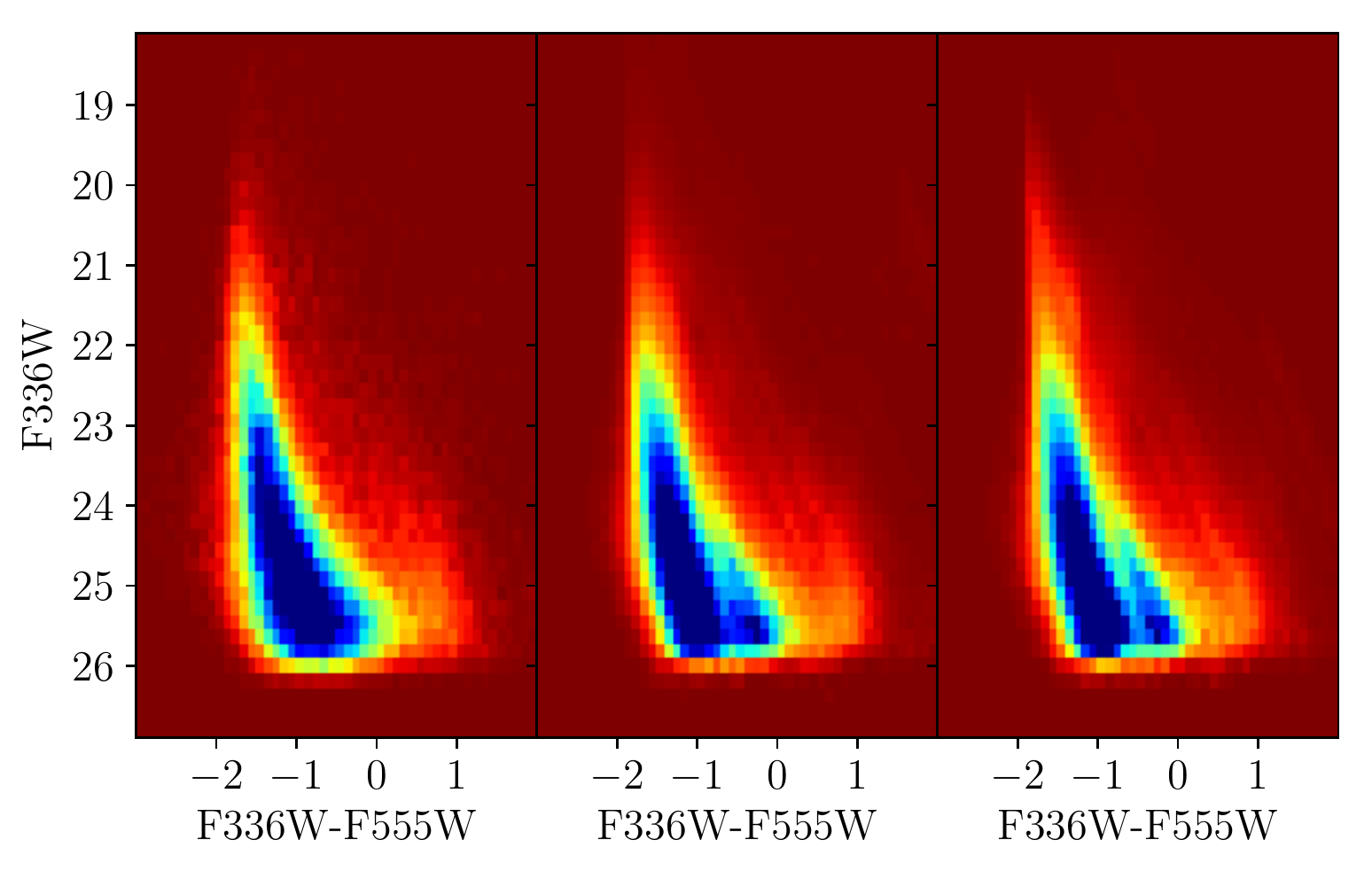}
\caption{Observational CMD of NGC~4449 (left panel) compared to the
  best synthetic CMDs (middle and right panel). Symbols are the
  same as in Fig.  \ref{3cmd_ugc4305}. }
\label{3cmd_ngc4449} 
\end{figure*}

\begin{figure}[t]
\centering \includegraphics[width=9cm]{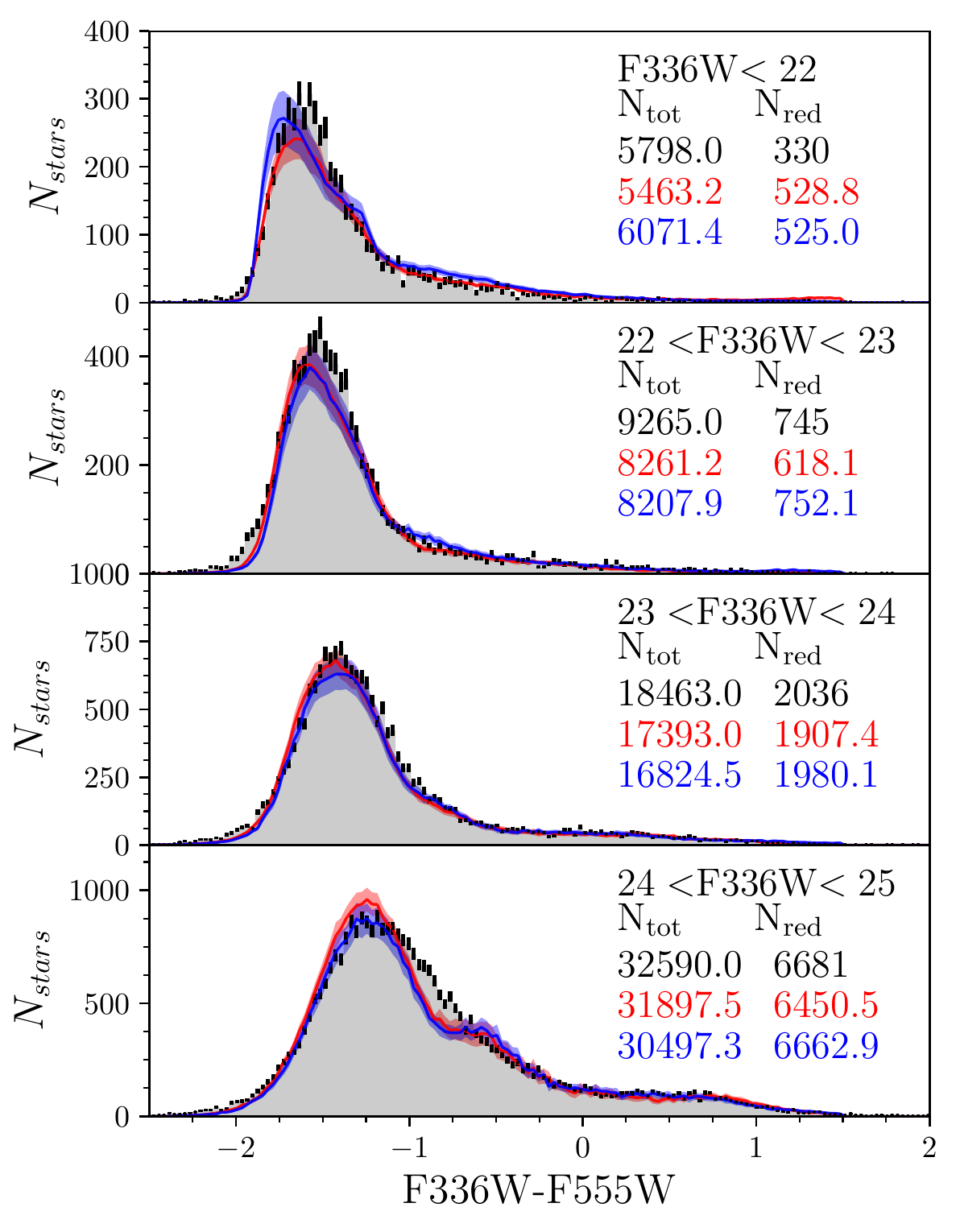}
\caption{{ Color distribution of NGC~4449 compared to the models
  computed with IMF exponent s$=-2.0$ above 5 M${_\odot}$. Symbols are
  the same as in Fig.  \ref{CF_ugc4305}.} }
\label{CF_ngc4449} 
\end{figure}

\begin{figure}[t]
\centering \includegraphics[width=9cm]{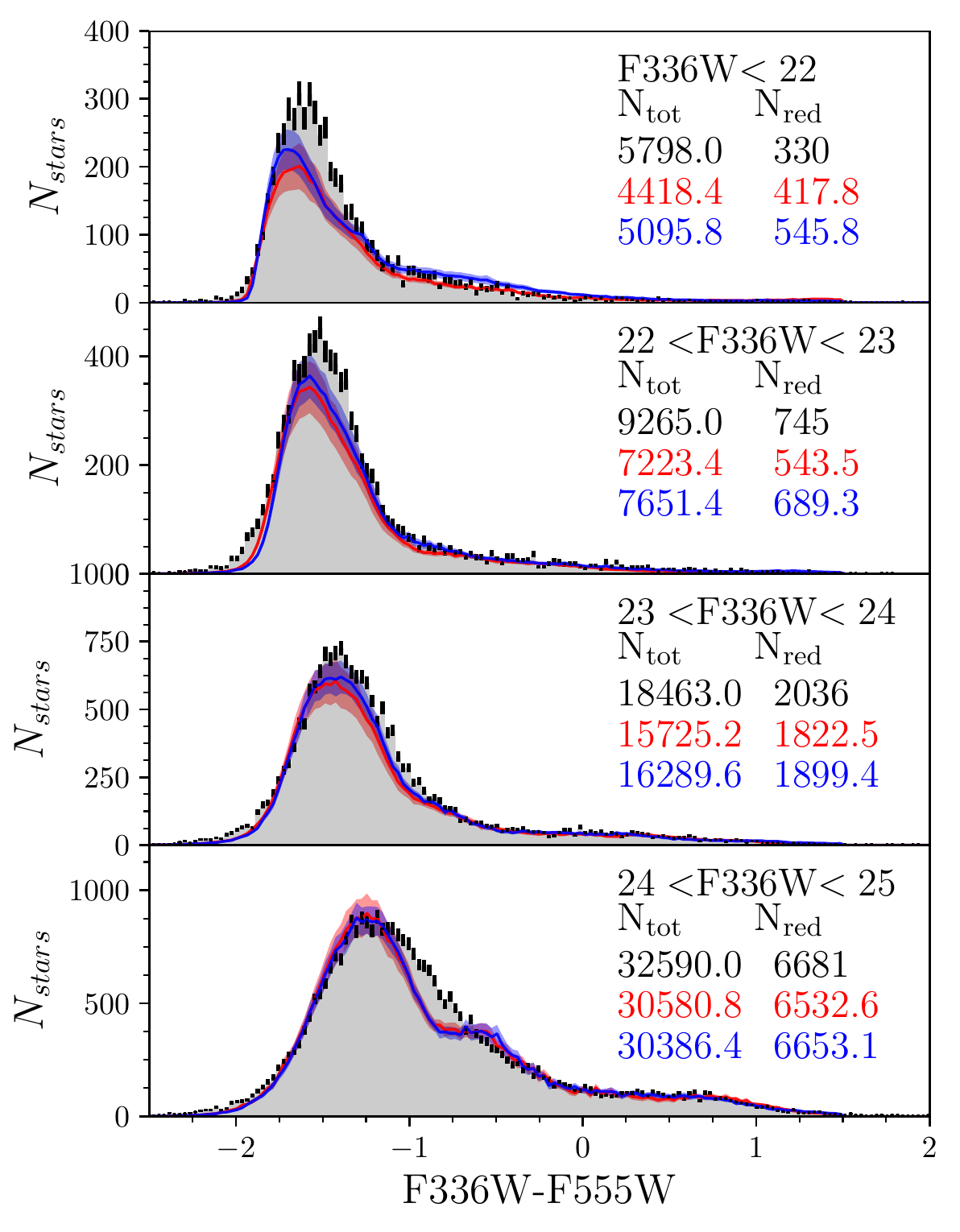}
\caption{{ Color distribution of NGC~4449 compared to the models
    computed with an IMF exponent s$=-2.3$ above 5
    M${_\odot}$. Symbols are the same as in Fig.  \ref{CF_ugc4305}.} }
\label{CF_ngc4449_salp} 
\end{figure}

{ The corresponding best synthetic CMDs and color distributions are
  shown in Fig. \ref{3cmd_ngc4449} and \ref{CF_ngc4449},
  respectively. For comparison, Fig. \ref{CF_ngc4449_salp} shows the
  color distribution using the canonical s$=-2.3$. Neither model is
    perfect. However, models with IMF exponent s$=-2.0$ provide a much
  better fit to the total number of stars brighter than $23$, with
  minimal differences below this limit. Moreover, irrespective of the
  adopted IMF, the number of post-MS stars (arbitrarily chosen with
  F336W$-$F555W$>-0.5$) is generally correct, except for F336W$<22$
  (see the top panel of Fig. \ref{CF_ngc4449} and
  \ref{CF_ngc4449_salp}) where we have an excess of synthetic stars
  (slightly exacerbated in the s$=-2.0$ case). Reducing further the
  IMF exponent has a different effect on the MIST solutions than
  on the PARSEC-COLIBRI ones. In the former the general likelihood of
  the synthetic models worsens, in the latter the likelihood slightly
  improves. Either way, the excess of synthetic post-MS stars is
  always exacerbated by a flatter IMF. A better fit could be obtained
  by varying the mass break-point where the IMF flattens, instead of
  using a single slope above 5 M${_{\odot}}$. Alternatively, the
  excess of objects with F336W$< 23$ could also indicate a high
  percentage of binaries with similar mass components (as found in
  starburst clusters, see e.g. \citealt{sana13}). Following mass
  accretion and mergers, close binary systems can create a tail of
  rejuvenated binary products, extending the single star mass function
  by up to a factor of 2 in mass (\citealt{schneider15}). Finally, we
  can not exclude that a fraction of compact, unresolved clusters could
  be contaminating the upper end of the CMD and luminosity function.

 Given these uncertainties and the strong
  degeneracy between mass and age for very massive stars, we refrain
  from further discussion of this issue. In terms of SFH, the main
  effect of adopting a slope $-2$ instead of $-2.3$ is to slightly
  decrease the recent SFR compared to the old one, hence making the
  galaxy SF slightly more constant.

 }


At fainter magnitudes (F336W$>24$, see the bottom panel of Figures
\ref{CF_ngc4449} and \ref{CF_ngc4449_salp}) we have another issue:
synthetic models show a star counts drop at colors around $-1$,
whereas the observed color distribution is instead continuous. {
  This result holds irrespective of the adopted IMF}. The presence of
this dip might be an indication that models are not able to produce
sufficiently extended loops during the HeB phase. Indeed, a similar dip was already
devised by \cite{tang14} in several dwarf galaxies. \cite{tang14}
showed that this discrepancy is overcome by extending the overshooting
at the base of the convective envelope. Another possibility is that
differential reddening dAV is underestimated. In fact: 1) the main
effect of differential reddening is to smear CMD features, like the
gap between MS and post-MS; 2) our estimate for dAV is based on the
assumption of flat reddening distribution (each synthetic star is
reddened with a random reddening between AV and AV$+$dAV), whereas the
real distribution might be more complex\footnote{We also tried a
  Gaussian reddening distribution with no significant fit
  improvement.} (see e.g. \citealt{harris97}) and mass/age
dependent. However, the color distribution of MS stars is generally
\emph{broader} in the model than in the data, suggesting that dAV is
not significantly overestimated.


{ The orange thick dashed} line in Fig. \ref{sfr_ngc4449} shows
the SFH derived by \cite{mq10} using the same optical ACS images
used in this work (\citealt{annibali08}). { In order to make a
  meaningful comparison, McQuinn's et al. SFRs were rescaled to account
  for the different spatial coverage. In fact, McQuinn's total FoV is
  more than twice as large as our FoV, being based on two partially
  overlapping ACS fields, whereas our UV observations consist of one
  single WFC3 field centered on the central part of the galaxy. The
  correction factor was estimated by counting the number of young
  stars inside and outside the WFC3 FoV in our optical CMDs (Sabbi et
  al. 2018). We found that only $\approx 8\%$ of the total recent SF
  in NGC~4449 was missed by our UV observations, hence McQuinn's
  solution was rescaled down by this amount. Since McQuinn's SFH
  assumes a Salpeter IMF between 0.1 and 120 M$_{\odot}$, we also show
  our extrapolated rates (thin blue and red dotted lines) using an
  exponent s$=-2.3$ between 0.1 M$_{\odot}$ and 5 M$_{\odot}$, and s
  $=-2.0$ above.

  Overall, our rates are within a factor of two of McQuinn's
  solution. Over the last 180 Myr, McQuinn et al. predict
  a continuous decrease of the SFR interrupted by a very recent and
  short SF enhancement between 4 (corresponding to the youngest
  isochrone in their models) and 10 Myr ago. Both our solutions follow
  closely this behavior, except for the epoch of the recent peak which
  is slightly older (10-16 Myr ago) in our solution. Part of the
  difference could be ascribed to the addition of the filter F336W,
  being more complete than the optical bands in crowded star forming
  regions (in the UV range, most cool stars appear very faint,
  resulting in less crowding). Other possible causes include: 1) the
  stellar models; those adopted here for intermediate and high mass
  stars are the latest on the market, whereas those adopted by McQuinn
  are the \cite{marigo07}, which rely on the older \cite{girardi00} for
  masses up to the 7 M$_{\odot}$, and \cite{bertelli94} above; 2) the
  different parameterization of the extinction; 3) the stellar species
  used in the fitting process; McQuinn's analysis is based on all
  stellar species including low mass stars ascending the RGB and
  TP-AGBs, ours is mostly based on high mass stars along the MS and
  BL.}

{ Our average rates (filled arrows on the right side of the plot)
  compare very well also with H$\alpha$ and FUV emission rates, as
  shown in Fig. \ref{sfr_ngc4449} with green and cyan thick dashed
  lines, respectively. As already done for Ho~II, the FUV SFR adopts the
  SFR$-$FUV calibration from \cite{k98} and the GALEX FUV flux,
  cropped to the footprint of the HST/ACS observations (taken from
  \citealt{mq15}) and corrected by 8\% to account the different
  spatial coverage, while the latter uses the ratio
  SFR(H$\alpha$)$/$SFR(FUV) provided by \cite{lee09}. Both FUV and
  H$\alpha$ rates assume a Salpeter IMF down to 0.1 M$_{\odot}$.}


An obvious question is what triggered the SF enhancement. If it is a
cold flow accretion or a merger event, nearby HI gas maps might show
morphologically disturbed features. Indeed, the global dynamics of the
HI associated with NGC~4449 show large distortions. A key
observational feature for this galaxy is neutral hydrogen extending to
6 times its Holmberg radius (see e.g. \citealt{vanw75, hunter98}),
characterised by an elongated ellipse of lower column density with a
major axis of 35 kpc. Beyond the ellipse, there is a complex
distribution of clouds and streamers that wrap around the galaxy. From
a kinematical point of view, the gas in the central 4 kpc diameter,
which corresponds to the brighter part of the optical galaxy, is seen
to exhibit rotation in the opposite direction to that of the ellipse
and streamers. Moreover, \cite{hunter99} found that the regions with
the highest velocity dispersions in HI are located outside the regions
of SF and in some cases beyond the bright part of the optical galaxy.

All these characteristics are consistent with a picture in which NGC
4449 has been disturbed by an external perturber, and the gas has not
yet returned to equilibrium. However, in contrast to the disturbed
morphology of the gas, NGC~4449 is a fairly isolated system. Its
closer neighbour, DDO~125, located at an apparent separation of 41 kpc
from the center of NGC 4449, does not show signs of damage and its
global gas and optical properties are typical of irregular galaxies
(\citealt{hunter97}). Between NGC~4449 and DDO~125 no clear bridge has
been detected. As discussed in \cite{hunter98}, with an HI mass of
only $10^8\,$M$_{\odot}$, it should have lost 90\% of its HI
mass in order to account for all of the mass in the streamers. On the other hand, \cite{mdg12} and \cite{rich12} detected and analyzed a
stellar tidal stream in the halo of NGC~4449, arguing that this object
is the result of the ongoing disruption of a dSph galaxy, while
\cite{annibali12} found a globular cluster of NGC~4449 which appears
associated with two tails of blue stars, possibly the nucleus of a
former gas-rich satellite galaxy undergoing tidal disruption by
NGC~4449. Finally, using metallicity measurements of the stream,
\cite{toloba16} suggested a progenitor as massive as Fornax or
Sagittarius. Interestingly, an interaction like that could have had a
substantial impact on NGC~4449.

\subsection{NGC~1705}

NGC~1705 is the most isolated of the three galaxies, since its closest
neighbour is at more than 500 kpc (LSBG F157-089).  Despite this, NGC
1705 has a warped HI disk (\citealt{meurer98}) that is significantly
off-set with respect to the stellar component, a possible signature
that the outer HI gas is still in the process of settling
down.

\begin{figure*}[t]
\centering \includegraphics[width=13cm]{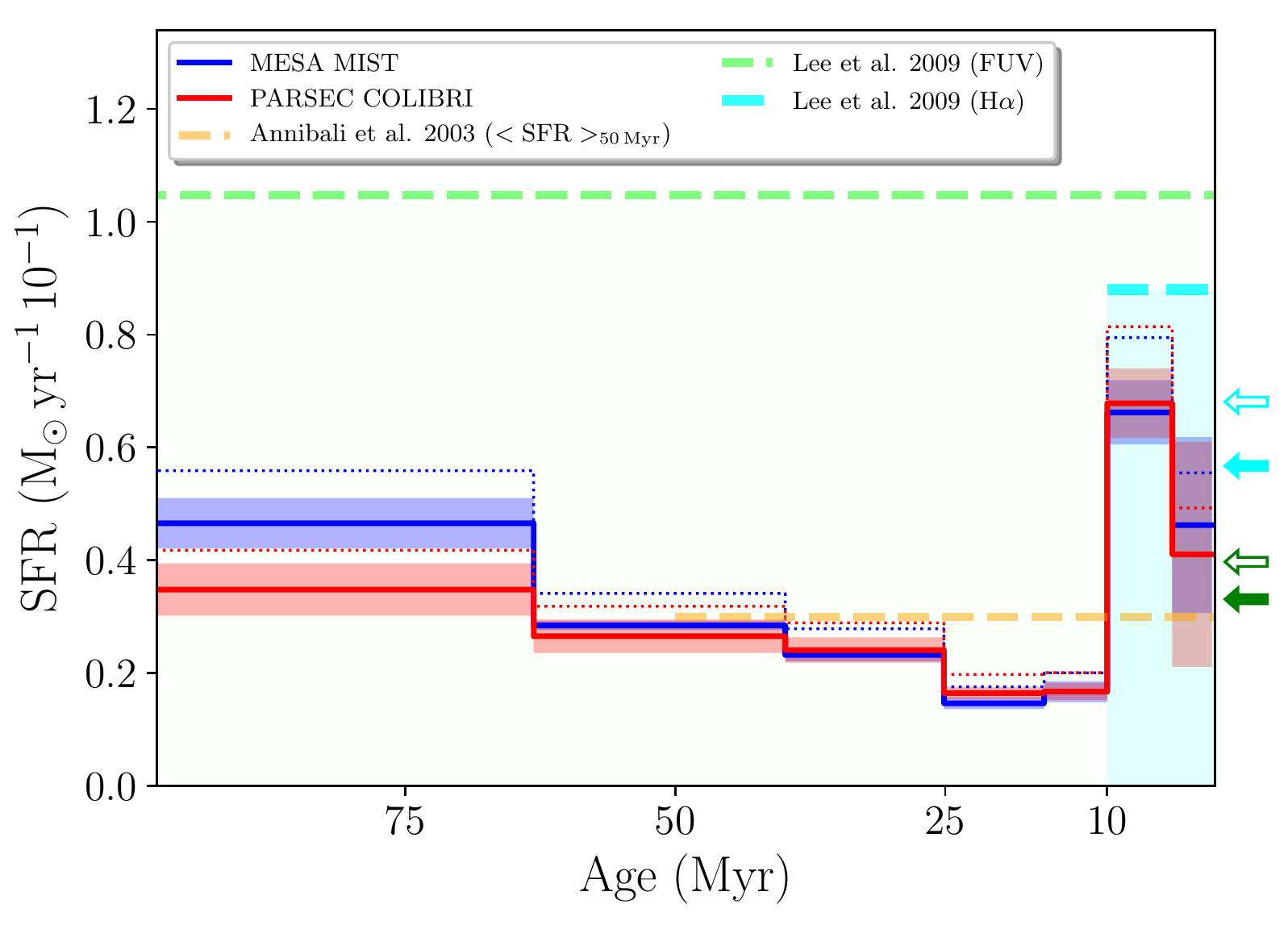}
\caption{Recovered SFH of NGC~1705. Symbols are the same as in
  Fig. \ref{sfr_ugc4305}. }
\label{sfr_ngc1705} 
\end{figure*}

Figure \ref{sfr_ngc1705} shows the recovered SFH for
NGC~1705. Overall, the activity of NGC~1705 has been slightly
declining for the past 100 Myr. This behaviour changed drastically 10
Myr ago, when the SFR increased by a factor of two or more over the
100 Myr-averaged-SFR. After the peak the SFR remained constant and no
drop is detected. The resulting extinction AV is $0.25$ mag, while the
differential one dAV is 0.12 mag. 

\begin{figure*}[t]
\centering \includegraphics[width=15cm]{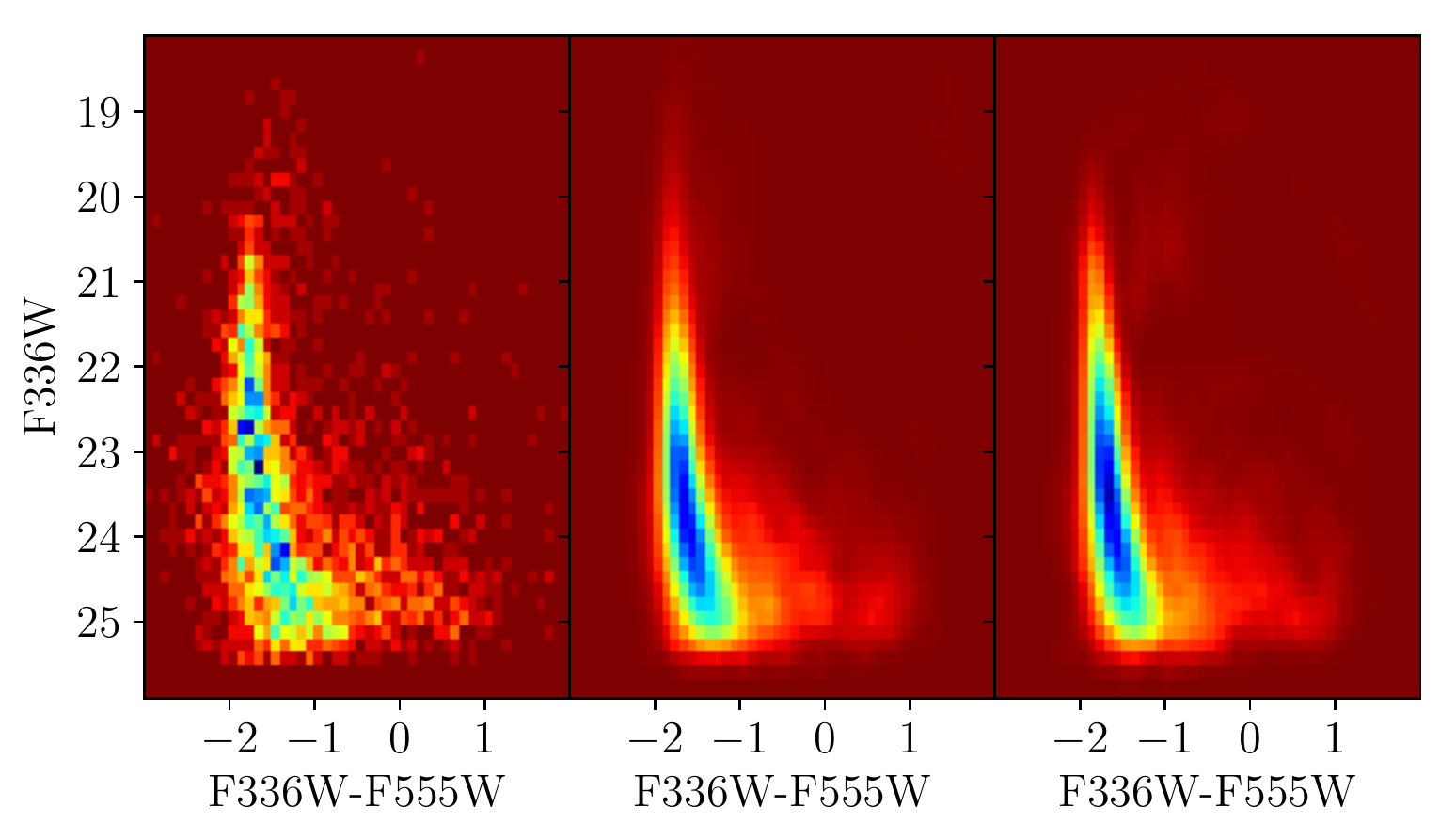}
\caption{Observational CMD of NGC~1705 (left panel) compared to the
  best synthetic CMDs (middle and right panel). Symbols are the
  same as in Fig.  \ref{3cmd_ugc4305}.}
\label{3cmd_ngc1705} 
\end{figure*}

\begin{figure}[t]
\centering \includegraphics[width=9cm]{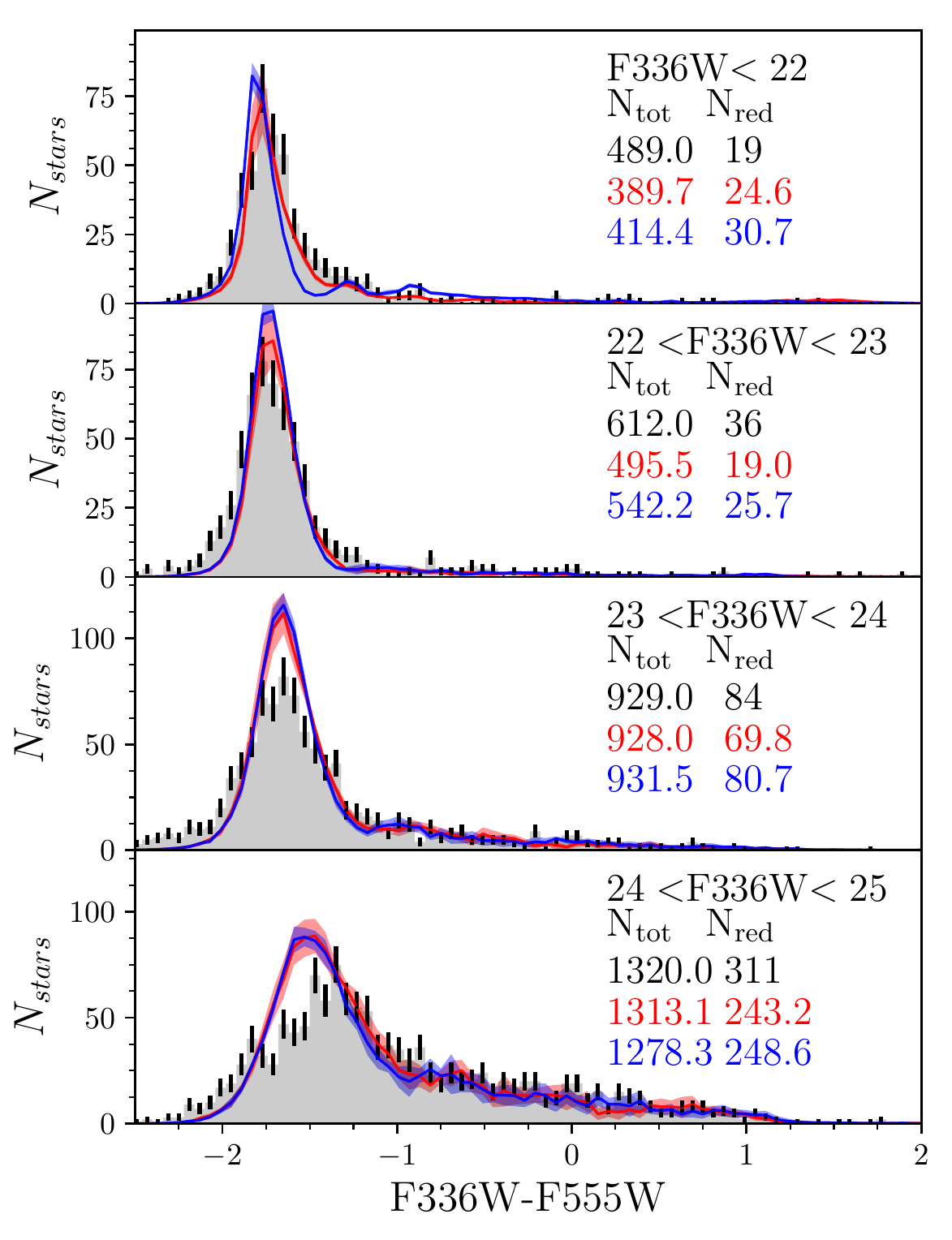}
\caption{Color distribution of NGC~1705 compared to the
  models. Symbols are the same as in Fig.  \ref{CF_ugc4305}.  }
\label{CF_ngc1705} 
\end{figure}

To visualize the quality of the reconstruction,
Fig. \ref{3cmd_ngc1705} shows the best synthetic CMD (middle and right
hand panels) compared to the observational counterpart, while
Fig. \ref{CF_ngc1705} shows the corresponding color distributions.  We
were unable to perfectly fit the upper end of the CMD, which is likely
populated by stars more massive than 40 M$_{\odot}$ (see
Fig. \ref{3cmd_iso}) { and appears to be broader than predicted by
  the synthetic CMDs. However, as for Ho~II, no modification of the
  assumed IMF exponent $-2.3$ is necessary.}  The synthetic CMDs show
a significant gap between the MS and the post-MS phase that instead is
completely absent in the observed CMD.  Interestingly, the gap holds
regardless of the rotational status of the stellar models
(PARSEC-COLIBRI are static and MIST are rotating).

This discrepancy is not surprising given the uncertainties in the
input physics of this mass range (see, e.g., \citealt{demink12}).
Significant mass-loss may cause a star to lose over half its mass
during its lifetime.  This mass loss occurs via stellar winds and
strongly affects the stellar evolution of a star. Moreover, massive
stars are often found to be rapid rotators and have a significant
preference to be binaries with orbital periods of less than a few days
(\citealt{sana11}).

Compared to the peak activity of NGC~4449 and Ho~II, the peak activity
of NGC~1705 is delayed by about 10 Myr. Moreover, whereas in NGC~4449 and
Ho~II the SFR dropped after the peak, the activity of NGC~1705 stayed
constant up to now. Our solution is qualitatively similar to the
findings of \cite{meurer92}, \cite{annibali03} and \cite{martin12},
although differences are found in the details.

\cite{meurer92} identifies three different populations: an unresolved
nucleus, whose activity is consistent with a SF burst 5-16 Myr old, an
inner high surface-brightness population, extending out to a radius of
500 pc and characterized by a mean age of 50 Myr and a young component
younger than 5 Myr (identified thanks to the presence of Wolf-Rayet
emission lines), and a low surface-brightness population at least 1.5
kpc wide, older than the previous two.

\cite{meurer92} studied the H${\alpha}$ emission line
kinematics of NGC 1705, detecting the presence of a kpc-scale
expanding super-shell of ionised gas centered on the central nucleus
with a blue-shifted emission component at 540 km s$^{-1}$.  Using an
expansion model, the authors find an expansion timescale of the order
of the age of the nucleus. This evidence, together with the
nucleocentric flow geometry, led the authors to conclude that the
nucleus itself is the likely the power source for the outflow.  UV
absorption line kinematics bears witness to this too
(\citealt{heck97,heck01}).

Based on HST/WFPC2 optical data, \cite{annibali03} studied the SFH
using the synthetic CMD approach. The authors identified two distinct
bursts. The older one (B1), confined to the most central regions and
occurred from 15 to 10 Myr ago and a stronger young burst (B2) started
$\approx 3$ Myr ago and still ongoing with a total SFR of
$\approx 0.3$ M$_{\odot}/$yr {(\citealt{annibali03} adopt a
  Salpeter IMF between 0.1 M$_{\odot}$ and 120 M$_{\odot}$)}. More
recently, \cite{annibali09} restudied the young populations of
NGC~1705 at shorter wavelengths (HST/WFPC2 in the F380W and F439W
bands, and HST/ACS/HRC in the F330W, F555W, and F814W broad-band
filters) confirming the presence of the younger burst B2. { The 50
  Myr-averaged-SFR of \cite{annibali03} is shown in Figure
  \ref{sfr_ngc1705} with a thick orange dashed line.}

{ Compared to Annibali's et al. solution, our SFH does not show any
  interruption over the past 10 Myr, and our main SF episode starts at
  an epoch intermediate between those of their two latest SF peaks,
  but the average rate is in good agreement. A possible origin for the
  different SFH shape could result from the adopted stellar models:}
Annibali et al. used the old Padova 1994 models that were available
for a few metallicities only, whereas our code incorporates the latest
PARSEC-COLIBRI and MIST calculations for a fine grid of
metallicity. Moreover, our WFC3 observations have a better resolution
than the WFPC2, and alleviate the problem of photometric blends.

{ Compared to integrated emission measurements, our 100
  Myr-averaged-SFR is about a factor of three lower than the rate
  derived by \cite{lee09} using FUV emission (green thick dashed line
  in Fig. \ref{sfr_ngc1705}), while our 10 Myr-averaged-SFR is only
  slightly lower than the rate derived by the same authors using
  H${\alpha}$ nebular emission (cyan thick dashed line in
  Fig. \ref{sfr_ngc1705}). However, these values are not resized to
  the WFC3 FoV, hence the excess of the FUV SFR may just indicate that
  the SF activity in the last 100 Myr is more widespread, extending
  well beyond the WFC3 FoV. Indeed, the much more spatially
  concentrated distribution of the recent and ongoing SF shows only a
  minor difference with the H${\alpha}$ rate. Moreover, part of the
  FUV and H${\alpha}$ emission could arise from unresolved clusters
  and associations, which are not considered in our rates.  }

Finally, our peak SF is in agreement with \cite{martin12}. This
confirms previous studies indicating that the core of NGC~1705 contains
massive stars, but no O and/or Wolf-Rayet objects. Using different
evolutionary tracks, they estimate the age of the super star cluster
to be $12 \pm 6$ Myr.

\section{Discussion}

A common feature in all recovered SFHs is the existence of a non-zero
(at 1 $\sigma$ level) SF activity at all ages. While the SFHs present
galaxy-to-galaxy variations, particularly at the youngest epochs, the
general trend is a relatively flat SF as a function of time. It is
important to point out, however, that time resolution at older ages
gets progressively worse, hence SF enhancement similar to those
detected in the last 30 Myr could be indistinguishable from a constant
activity 100$-$200 Myr ago.

\begin{figure}[t] 
\centering \includegraphics[width=9cm]{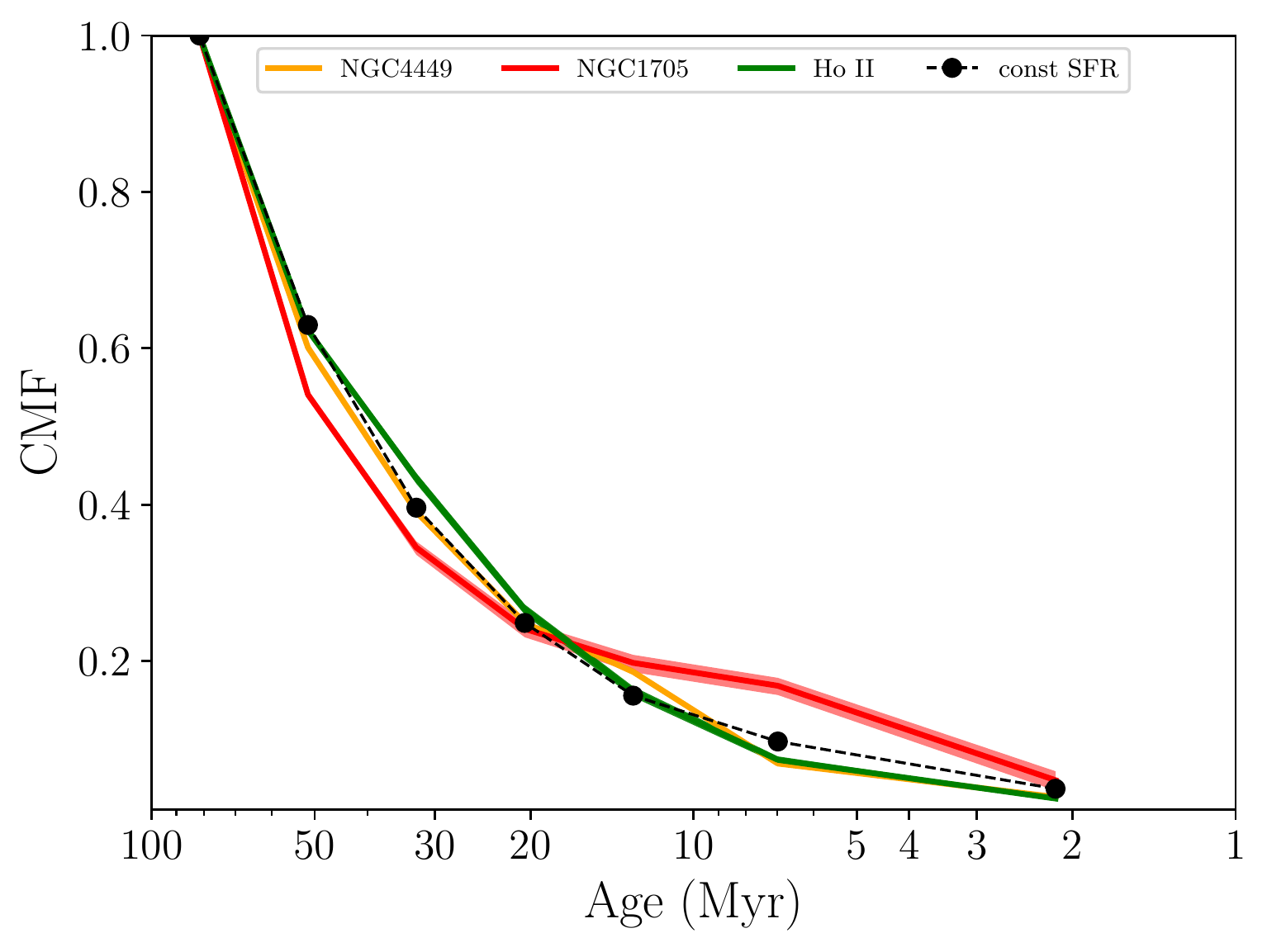}
\caption{Cumulative stellar mass fraction as a function of time for the three dwarfs and for a constant SFR. }
\label{cumul_SFH} 
\end{figure}

All the three dwarfs show SF enhancements in the last 50 Myr, but the
SF is not always continuing to the present time at a very high
rate. The strongest enhancement is found in NGC~4449, which is
followed by a lower activity in the last 10 Myr. The situation is
opposite in NGC~1705, where the highest rate in the last 50 Myr is
reached in the \emph{last} 10 Myr. Ho~II is qualitatively similar to
NGC~4449, even though the ratio between the peak activity and the
following drop is much lower. These differences are promptly evident
when one compares the cumulative stellar mass fraction (CMF) for the
three dwarfs (Fig. \ref{cumul_SFH}) in the last 100 Myr (NGC~1705's
solution becomes too uncertain at older epochs). The CMF, i.e., the
fraction of the total stellar mass formed at a given time, provides a
normalized measure of the stellar mass accumulation. Compared to the
absolute SFHs, cumulative ones are less affected by correlated SFRs in
adjacent time bins.  In order to combine the PARSEC-COLIBRI and MIST
solutions, the CMFs of Fig. \ref{cumul_SFH} are the weighted
mean\footnote{If we consider the two solutions as two normal independent
  variables with means z1 and z2, and dispersions $\sigma_1$ and
  $\sigma_2$, the combined solution is normal with mean
  $({\sigma_1}^{-2}*z1+{\sigma_2}^{-2}*z2)/({\sigma_1}^{-2}+{\sigma_2}^{-2})$
  and dispersion $(1/({\sigma_1}^{-2}+{\sigma_2}^{-2}))^{-1/2}$. } of
the two for each galaxy.

The net result is that NGC~1705 systematically deviates from NGC~4449
and Ho~II. Using a constant mass production (dashed line) as a
reference model, while NGC~1705's production is clearly above the
reference model in the last 10 Myr (and below the reference model
between 30 and 50 Myr), NGC~4449 shows a relative excess of mass
between 10 and 16 Myr ago. On the other hand Ho~II is almost
indistinguishable from the constant mass production.

{ At this point an interesting question is whether or not the most
  recent peak is a unique event in the recent history of our
  galaxies. Simulations taking into account feedback processes
  illustrate that star formation in isolated dwarf galaxies can
  exhibit episodic or cyclic behaviour
  (e.g. \citealt{stinson07}). Here, the starburst is terminated by the
  expulsion of gas through supernova winds. If the gas later accretes
  back to the galaxy, a new starburst could be ignited.  Concerning
  our sample, the answer depends on the look-back time. In the last 16
  Myr, NGC~4449 assembled a stellar mass between $7.7\times 10^{6}$
  M$_{\odot}$ and $9.0\times 10^{6}$ M$_{\odot}$. If we dilute this
  mass between 25 and 40 Myr ago, the resulting SFR ($0.52-0.60$
  M$_{\odot}/$yr) would be \emph{higher} than the measured rate
  ($0.40-0.46$ M$_{\odot}/$yr), whereas between 40 and 63 Myr ago the
  resulting SFR would be \emph{lower} than the measured rate (hence we
  can not exclude multiple events like the current one or
  stronger). Taking these numbers at face value and considering the
  error bars of our SFH, we find unlikely that another SF enhancement
  similar to the most recent one has taken place in NGC~4449 in the
  past 40 Myr, whereas at older epochs the situation is much more
  uncertain. In other words, the recent enhancement of NGC~4449 is
  probably a unique event in the last 40 Myr, but nothing can be said
  for the earlier epochs. From this point of view, a SFH composed by
  periodic bursts with duration 5 Myr, period 40 Myr and peak rate
  like the current one, would be indistinguishable from a constant
  activity prior to 40 Myr ago. From a dynamical point of view, using
  N-Body simulations, \cite{theis99} found that the closest approach
  between NGC~4449 and the nearby dwarf DDO~125 happened 350 Myr ago
  at a minimum distance of 25 kpc, while \cite{rich12} found a
  timescale of the order of $10^8$ yr for the pericentric passage of
  NGC~4449B, a dwarf galaxy companion of NGC~4449 in a transient stage
  of tidal disruption.

  In a forthcoming paper, Sacchi et al. 2018, we will explore these
  older epochs using deeper HST/ACS optical images.

A similar conclusion can be reached for NGC~1705. In the last 10 Myr
this galaxy assembled about $5.7\times 10^{5}$ M$_{\odot}$ of stars. If a
similar stellar mass is diluted between 25 and 40 Myr ago, the
resulting SFR would be higher than the measured rate, whereas at older
epochs it would be comparable (hence suggesting that at these epochs
another SF event similar to the current one can not be ruled out).}

Concerning Ho~II, the combination of large statistical uncertainties
and lower peak rate prevents any conclusion about the presence of
previous peaks similar to the current one.

\begin{figure}[t]
\centering \includegraphics[width=9cm]{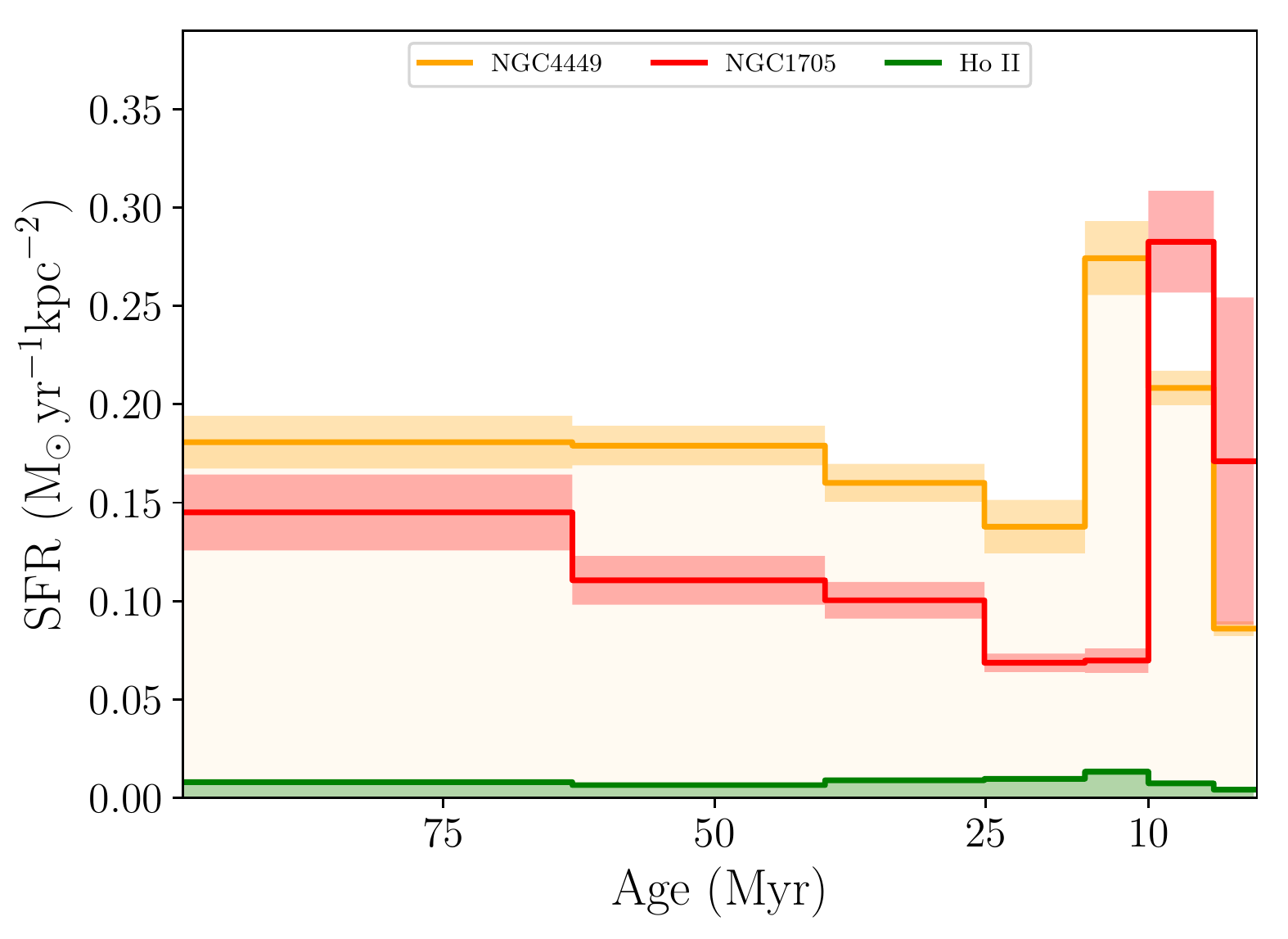}
\caption{SFR density per unit area for NGC~4449 (gold line), NGC~1705
  (red line) and Ho~II (green line). { For each galaxy the adopted
    area is that of the region containing 90$\%$ of the stars in our
    catalogue.}  Only PARSEC-COLIBRI solutions are shown.}
\label{dens_SFH} 
\end{figure}

{ Finally, we must consider the effect of size, since NGC~1705 is
  significantly more compact than NGC~4449 and Ho~II.  We thus compare
  the SFRs per unit area. Since our photometry doesn't cover the
  entire extent of the galaxies, we consider the SFR density only over
  a meaningful portion of our dwarfs, and we have chosen the region in
  each galaxy whose isodensity contour encloses 90\% of all the stars
  in our catalogue. } The corresponding area is used to normalize the
SFHs, as shown in Fig.  \ref{dens_SFH}. Of course this is a crude
estimate since neither the photometric completeness is the same nor
different generations have the same spatial distribution. In general,
the more concentrated the stars within the galaxy, the more
underestimated the star formation density. Interestingly, despite this
bias (which disfavors NGC~1705), Figure \ref{dens_SFH} clearly shows
that the projected star formation density peak of NGC~1705 is
comparable with that of NGC~4449, and one order of magnitude higher
than that of Ho~II. 


\section{Conclusions}

{ The main goal of this paper was a detailed analysis of the UV
  CMDs of the younger resolved stars in three well studied active
  dwarf galaxies, NGC~4449, NGC~1705 and Ho~II, not only in terms of
  quantitative SFHs but also in terms of reliability of the most
  popular stellar evolution models for intermediate and high mass
  stars.  For this task we have compared deep HST F336W vs
  F336W$-$F555W CMDs with synthetic ones generated with
  state-of-the-art stellar models, namely PARSEC-COLIBRI and
  MIST. This has allowed us to study the recent history of these
  galaxies with unprecedented resolution, emphasizing the strengths
  and weaknesses of these new models. Our main results are summarized
  as follows.}

\begin{itemize}

\item { In general, we found a good overall agreement between the
    SFHs derived using PARSEC-COLIBRI and MIST. However, significant
    systematic deviations emerged as well, suggesting that the
    physical assumptions of micro- and macro-physics have still an
    impact on the recovered rates. Indeed, this result is
    not surprising given the clear morphological differences already
    visible in the synthetic CMDs of Fig. \ref{6cmd}, with both color
    and luminosities affected. This result is especially important if
    one considers that PARSEC-COLIBRI or MIST only represent two
    possible realistic parameterizations, whereas a full estimate of
    the real uncertainties should take into account different
    assumptions about critical parameters like stellar mass loss,
    overshooting, rotation and binary evolution. From this point of
    view, the differences between PARSEC-COLIBRI and MIST SFHs should
    be taken as lower limits to the real uncertainties due to stellar evolution modeling. }

  { Concerning the SFHs}, no strong SF burst was detected in the
  last 180 Myr in any of the three starburst dwarfs. The SFR
  increases, at most, by a factor of $\approx 2$ over the 100
  Myr-averaged-SFR. This ``flickering'' SF mode is typically seen in
  the last 100 Myr of many high resolution studies (see
  e.g. \citealt{weisz08,mq10}) and is probably connected with the
  formation of star clusters and associations. Multiple flickering
  events could then form the long-lasting ($>\,450$ Myr) bursts
  detected by \cite{mq10} in several dwarfs, whose origin is probably
  global. { Compared with other independent SFR tracers, our 100
    Myr-averaged-SFR and 10 Myr-averaged-SFR are also in good
    agreement with FUV and H${\alpha}$ rates, respectively.}
  
  Our three dwarfs show qualitatively similar histories: a recent
  major SF enhancement superimposed on a rather constant SF
  activity. Quantitatively, instead, the SFRs differ from each other
  by orders of magnitude and the details of the most recent peak are
  remarkably different.  The recent SFR in Ho~II and NGC~4449 peaked
  10$-$15 Myr ago and has been declining since then, whereas the SF activity in
  NGC~1705 peaked 10 Myr ago and is still ongoing. Moreover, once
  normalized to the area, NGC~1705 activity is comparable to that of
  the far more massive NGC~4449.

  { Concerning the timing of the most recent peak, the SF
    enhancements of NGC~4449 and NGC~1705 are probably the only major
    SF event in the history of these galaxies over the last 40 Myr. A
    similar conclusion can not be drawn for Ho~II, since the peak
    activity is too low to be detected even 40 Myr ago with current
    data.}

  Finally, the truly isolated nature of NGC~1705 compared to NGC~4449
  and Ho~II suggests that close interactions with neighbouring
  galaxies may not be the only explanation for the recent SF
  enhancement. Indeed, the clearly disturbed gas kinematics (see
  e.g. \citealt{lelli14}) of NGC~1705 could point to a cold accretion
  of gas as the responsible process for the current SF event.

\item { In terms of IMF, a Salpeter-like power law down to 5
    M$_{\odot}$ is consistent with the data of Ho~II and NGC~1705,
    whereas in NGC~4449 a flatter IMF (s$=-2.0$) provides a better fit
    to the data. A similar result has been recently found for the
    starburst region of 30 Doradus (\citealt{schneider18}). Using
    spectroscopy, these authors found a $32 \pm 12$\% more stars above
    30 M$_{\odot}$ than predicted by a standard Salpeter
    IMF. Interestingly, in NGC~4449 the excess of stars (depending on
    the model, between 15 and 23 \% with respect to Salpeter) is for
    magnitudes brighter than F336W$\approx 23$, where at least 50\% of
    the stars are more massive than 20 M$_{\odot}$.  }

\item While all models generally reproduce the expected numbers of MS
  and post-MS stars in different magnitude bins, the colour
  distributions show several issues which cannot be readily fixed by
  changing metallicity. The transition from MS and post-MS is
  especially problematic in NGC~4449, the only one with a
  statistically well populated CMD. { At the bright end the total
    number of massive stars can be reproduced using a flatter IMF},
  but not the ratio between MS and post-MS, with a clear excess of the
  latter. This behavior is further complicated at the faint end of
  NGC~4449 (F336W$>$23), where all models predict a gap between MS and
  post-MS around F336W$-$F555W$\approx -1$, whereas the observational
  CMD shows a smooth transition. We also found that NGC~4449 is
  affected by a significant differential reddening.

  The best synthetic models for Ho~II do not show significant
  discrepancies with the observational counterpart. Interestingly,
  this galaxy has negligible differential reddening and lower
  metallicity compared to NGC~4449, suggesting that both these factors
  may have a role behind the mismatches of NGC~4449. On the
  metallicity side, this may suggest a difficulty of models in
  matching the BL length at higher metallicity. In this direction, the
  best fitting CMDs of the similarly metal rich NGC~1705 show
  analogous issues. The most massive stars of NGC~1705 make a smooth
  transition from MS to post-MS, with no clear distinction between the
  two species, whereas all models predict a net gap.

\end{itemize}

{ We wish to emphasize once again that the above results have been
  obtained thanks to the exquisite color and time resolution allowed
  by UV photometry. While optical CMDs have the advantage of reaching
  much older look-back times, thus allowing to derive the SFH back to
  much earlier epochs, UV CMDs allow us to separate much better MS and
  post-MS stars, thus distinguishing in more detail the most recent
  history. To these two complementary viewpoints, in the future it
  will be crucial to add also the NIR one, when JWST will allow to
  study in detail the reddest evolutionary phases, from the youngest
  ones of stars still embedded in their parent cocoon to the latest
  AGB and thermally pulsating stars, to the old Red Clump ones.  }
  
\acknowledgments 

{ We warmly thank Kristy McQuinn for her helpful discussions and
  suggestions.  M.C. and M.T.  acknowledge funding from the INAF
  PRIN-SKA 2017 program 1.05.01.88.04.}  D.A.G. kindly acknowledges
financial support by the German Research Foundation (DFG) through
programme GO 1659/3-2. B.G.E. acknowledges support from NASA grant
HST-GO-13364.14-A.

\bibliographystyle{apalike}

\begin{thebibliography}{}

\bibitem[Annibali et al.(2003)]{annibali03} Annibali, F., Greggio, L., Tosi, M., Aloisi, A., \& Leitherer, C.\ 2003, \aj, 126, 2752 

\bibitem[Annibali et al.(2008)]{annibali08} Annibali, F., Aloisi, A.,
  Mack, J., et al.\ 2008, \aj, 135, 1900 

\bibitem[Annibali et al.(2009)]{annibali09} Annibali, F., Tosi, M., Monelli, M., et al.\ 2009, \aj, 138, 169 


\bibitem[Annibali et al.(2012)]{annibali12} Annibali, F., Tosi, M., Aloisi, A., van der Marel, R.~P., \& Martinez-Delgado, D.\ 2012, \apjl, 745, L1 


\bibitem[Annibali et al.(2015)]{annibali15} Annibali, F., Tosi, M., Pasquali, A., et al.\ 2015, \aj, 150, 143 

\bibitem[Annibali et al.(2016)]{annibali16} Annibali, F., Nipoti, C., Ciotti, L., et al.\ 2016, \apjl, 826, L27 

\bibitem[Bellazzini et al.(2001)]{bellazzini01} Bellazzini, M., Ferraro, F.~R., \& Pancino, E.\ 2001, \apj, 556, 635 

\bibitem[Bekki \& Couch(2003)]{bekki03} Bekki, K., \& Couch, W.~J.\ 2003, \apjl, 596, L13 

\bibitem[Bekki \& Freeman(2002)]{bekki02} Bekki, K., \& Freeman, K.~C.\ 2002, \apjl, 574, L21 

\bibitem[Berg et al.(2012)]{berg12} Berg, D.~A., Skillman, E.~D., Marble, A.~R., et al.\ 2012, \apj, 754, 98 



\bibitem[Bernard et al.(2012)]{bernard12} Bernard, E.~J., Ferguson, A.~M.~N., Barker, M.~K., et al.\ 2012, \mnras, 426, 3490 



\bibitem[Bertelli et al.(1994)]{bertelli94} Bertelli, G., Bressan, A.,
  Chiosi, C., Fagotto, F., \& Nasi, E.\ 1994, \aaps, 106,

\bibitem[Bressan et al.(2012)]{bressan12} Bressan, A., Marigo, 
P., Girardi, L., et al.\ 2012, \mnras, 427, 127 


\bibitem[Bressan et al.(2015)]{bressan15} Bressan, A., Girardi, L., Marigo, P., Rosenfield, P., \& Tang, J.\ 2015, Asteroseismology of Stellar Populations in the Milky Way, 39, 25 



\bibitem[Bureau \& Carignan(2002)]{BC02} Bureau, M., \& Carignan, C.\ 2002, \aj, 123, 1316 

\bibitem[Calzetti et al.(2015)]{calzetti15} Calzetti, D., Lee, J.~C., Sabbi, E., et al.\ 2015, \aj, 149, 51 


\bibitem[Caffau et al.(2011)]{caffau11} Caffau, E., Ludwig, H.-G., Steffen, M., Freytag, B., \& Bonifacio, P.\ 2011, \solphys, 268, 255 

\bibitem[Cardelli et al.(1989)]{cardelli89} Cardelli, J.~A., Clayton,
  G.~C., \& Mathis, J.~S.\ 1989, \apj, 345, 245 

\bibitem[Choi et al.(2016)]{choi16} Choi, J., Dotter, A., Conroy, C., et al.\ 2016, \apj, 823, 102 


\bibitem[Cignoni et al.(2015)]{cigno15} Cignoni, M., Sabbi, E., van der Marel, R.~P., et al.\ 2015, \apj, 811, 76 

\bibitem[Cignoni et al.(2016)]{cigno16} Cignoni, M., Sabbi, E., van der Marel, R.~P., et al.\ 2016, \apj, 833, 154 

\bibitem[Croxall et al.(2009)]{crox09} Croxall, K.~V., van Zee, L., Lee, H., et al.\ 2009, \apj, 705, 723-738 

\bibitem[Dalcanton et al.(2012)]{dalcanton12} Dalcanton, J.~J., Williams, B.~F., Melbourne, J.~L., et al.\ 2012, \apjs, 198, 6 

\bibitem[Firth et al.(2006)]{firth06} Firth, P., Evstigneeva, E.~A., Jones, J.~B., et al.\ 2006, \mnras, 372, 1856 

\bibitem[de Mink et al.(2012)]{demink12} de Mink, S.~E., Brott, I., Cantiello, M., et al.\ 2012, Proceedings of a Scientific Meeting in Honor of Anthony F.~J.~Moffat, 465, 65 


\bibitem[Dohm-Palmer et al.(1997)]{dp} Dohm-Palmer, R.~C., Skillman, E.~D., Saha, A., et al.\ 1997, \aj, 114, 2527 

\bibitem[Dotter(2016)]{dotter16} Dotter, A.\ 2016, \apjs, 222, 8 


\bibitem[Evstigneeva et al.(2007)]{evsti07} Evstigneeva, E.~A., Drinkwater, M.~J., Jurek, R., et al.\ 2007, \mnras, 378, 1036 

\bibitem[Elmegreen et al.(2012)]{elme12} Elmegreen, B.~G., Zhang, H.-X., \& Hunter, D.~A.\ 2012, \apj, 747, 105 

\bibitem[Gallart et al.(2015)]{gallart15} Gallart, C., Monelli, M., Mayer, L., et al.\ 2015, \apjl, 811, L18 

\bibitem[Gerola et al.(1980)]{gerola80} Gerola, H., Seiden, P.~E., \& Schulman, L.~S.\ 1980, \apj, 242, 517 

\bibitem[Girardi et al.(2000)]{girardi00} Girardi, L., Bressan, A., Bertelli, G., \& Chiosi, C.\ 2000, \aaps, 141, 371 

\bibitem[Harris et al.(1997)]{harris97} Harris, J., Zaritsky, D., \&
  Thompson, I.\ 1997, \aj, 114, 1933 

\bibitem[Heckman \& Leitherer(1997)]{heck97} Heckman, T.~M., \& Leitherer, C.\ 1997, \aj, 114, 69 

\bibitem[Heckman et al.(2001)]{heck01} Heckman, T.~M., Sembach, K.~R., Meurer, G.~R., et al.\ 2001, \apj, 554, 1021 

\bibitem[Hill et al.(1994)]{hill94} Hill, R.~S., Home, A.~T., Smith, A.~M., et al.\ 1994, \apj, 430, 568 

\bibitem[Hunter et al.(1999)]{hunter99} Hunter, D.~A., van Woerden, H., \& Gallagher, J.~S.\ 1999, \aj, 118, 2184 


\bibitem[Hunter(1997)]{hunter97} Hunter, D.\ 1997, \pasp, 109, 937 



\bibitem[Hunter et al.(1998)]{hunter98} Hunter, D.~A., Wilcots, E.~M., van Woerden, H., Gallagher, J.~S., \& Kohle, S.\ 1998, \apjl, 495, L47 

\bibitem[Hunter \& Elmegreen(2004)]{he04} Hunter, D.~A., \& Elmegreen, B.~G.\ 2004, \aj, 128, 2170 

\bibitem[Huchtmeier \& Richter(1989)]{hr89} Huchtmeier, W.~K., \& Richter, O.~G.\ 1989, Science, 246, 943 


\bibitem[Karachentsev et al.(2002)]{kara02} Karachentsev, I.~D., Dolphin, A.~E., Geisler, D., et al.\ 2002, \aap, 383, 125 

\bibitem[Karczewski et al.(2013)]{kar13} Karczewski, O.~{\L}., Barlow, M.~J., Page, M.~J., et al.\ 2013, \mnras, 431, 2493 

\bibitem[Kennicutt(1998)]{k98} Kennicutt, R.~C., Jr.\ 1998, \araa, 36, 189 

\bibitem[Koleva et al.(2014)]{koleva14} Koleva, M., De Rijcke, S., Zeilinger, W.~W., et al.\ 2014, \mnras, 441, 452 

\bibitem[Kere{\v s} et al.(2005)]{keres15} Kere{\v s}, D., Katz, N., Weinberg, D.~H., \& Dav{\'e}, R.\ 2005, \mnras, 363, 2 

\bibitem[Kroupa(2001)]{kroupa01} Kroupa, P.\ 2001, \mnras, 322, 
231 

\bibitem[Lee et al.(2009)]{lee09} Lee, J.~C., Gil de Paz, A.,
  Tremonti, C., et al.\ 2009, \apj, 706, 599-613 

\bibitem[Lelli et al.(2012)]{lelli12} Lelli, F., Verheijen, M., Fraternali, F., \& Sancisi, R.\ 2012, \aap, 544, A145 

\bibitem[Lelli et al.(2014)]{lelli14} Lelli, F., Verheijen, M., \&
  Fraternali, F.\ 2014, \mnras, 445, 1694 

\bibitem[Marigo \& Girardi(2007)]{marigo07} Marigo, P., \& Girardi, L.\ 2007, \aap, 469, 239 

\bibitem[Marigo et al.(2017)]{marigo17} Marigo, P., Girardi, L., Bressan, A., et al.\ 2017, \apj, 835, 77 

\bibitem[Mart{\'{\i}}nez-Delgado et al.(2012)]{mdg12} Mart{\'{\i}}nez-Delgado, D., Romanowsky, A.~J., Gabany, R.~J., et al.\ 2012, \apjl, 748, L24 

\bibitem[Martins et al.(2012)]{martin12} Martins, F., F{\"o}rster Schreiber, N.~M., Eisenhauer, F., \& Lutz, D.\ 2012, \aap, 547, A17 




\bibitem[McQuinn et al.(2010)]{mq10} McQuinn, K.~B.~W., Skillman,
  E.~D., Cannon, J.~M., et al.\ 2010, \apj, 724, 49 

\bibitem[McQuinn et al.(2011)]{mq11} McQuinn, K.~B.~W., Skillman, E.~D., Dalcanton, J.~J., et al.\ 2011, \apj, 740, 48 

\bibitem[McQuinn et al.(2012)]{mq12} McQuinn, K.~B.~W., Skillman, E.~D., Dalcanton, J.~J., et al.\ 2012, \apj, 759, 77 

\bibitem[McQuinn et al.(2015)]{mq15} McQuinn, K.~B.~W., Skillman, E.~D., Dolphin, A.~E., \& Mitchell, N.~P.\ 2015, \apj, 808, 109 



\bibitem[Meynet \& Maeder(1997)]{mm97} Meynet, G., \& Maeder, A.\ 1997, \aap, 321, 465 



\bibitem[Melisse \& Israel(1994)]{mi94} Melisse, J.~P.~M., \& Israel, F.~P.\ 1994, \aap, 285, 51 


\bibitem[Meurer et al.(1992)]{meurer92} Meurer, G.~R., Freeman, K.~C., Dopita, M.~A., \& Cacciari, C.\ 1992, \aj, 103, 60 

\bibitem[Meurer et al.(1998)]{meurer98} Meurer, G.~R., Staveley-Smith, L., \& Killeen, N.~E.~B.\ 1998, \mnras, 300, 705 


\bibitem[Moustakas et al.(2010)]{mou10} Moustakas, J., Kennicutt, R.~C., Jr., Tremonti, C.~A., et al.\ 2010, \apjs, 190, 233-266 


\bibitem[Patton et al.(2013)]{patton13} Patton, D.~R., Torrey, P., Ellison, S.~L., Mendel, J.~T., \& Scudder, J.~M.\ 2013, \mnras, 433, L59 

\bibitem[Paxton et al.(2011)]{paxton11} Paxton, B., Bildsten, L., Dotter, A., et al.\ 2011, \apjs, 192, 3 

\bibitem[Paxton et al.(2013)]{paxton13} Paxton, B., Cantiello, M., Arras, P., et al.\ 2013, \apjs, 208, 4 

\bibitem[Paxton et al.(2015)]{paxton15} Paxton, B., Marchant, P., Schwab, J., et al.\ 2015, \apjs, 220, 15 

\bibitem[Puche et al.(1992)]{puche92} Puche, D., Westpfahl, D., Brinks, E., \& Roy, J.-R.\ 1992, \aj, 103, 1841 

\bibitem[Ramya et al.(2009)]{ramya09} Ramya, S., Kantharia, N.~G., \& Prabhu, T.~P.\ 2009, The Low-Frequency Radio Universe, 407, 114 

\bibitem[Rich et al.(2012)]{rich12} Rich, R.~M., Collins, M.~L.~M., Black, C.~M., et al.\ 2012, \nat, 482, 192 

\bibitem[Sabbi et al.(2018)]{sabbi18} Sabbi, E. and LEGUS collaboration, \apjs, accepted


\bibitem[Sacchi et al.(2016)]{sacchi16} Sacchi, E., Annibali, F.,
  Cignoni, M., et al.\ 2016, \apj, 830, 3 

\bibitem[Sacchi et al.(2017)]{sacchi17} Sacchi, E., et al., 2017, ApJ, submitted  

\bibitem[Sana et al.(2013)]{sana13} Sana, H., de Koter, A., de Mink, S.~E., et al.\ 2013, \aap, 550, A107 

\bibitem[Sana \& Evans(2011)]{sana11} Sana, H., \& Evans, C.~J.\ 2011, Active OB Stars: Structure, Evolution, Mass Loss, and Critical Limits, 272, 474 




\bibitem[Schaerer et 
al.(1993)]{schaerer93} Schaerer, D., Meynet, G., Maeder, A., \& Schaller, G.\ 1993, \aaps, 98, 523 


\bibitem[Schneider et al.(2015)]{schneider15} Schneider, F.~R.~N.,
  Izzard, R.~G., Langer, N., \& de Mink, S.~E.\ 2015, \apj, 805, 20 

\bibitem[Schneider et al.(2018)]{schneider18} Schneider, F.~R.~N., Sana, H., Evans, C.~J., et al.\ 2018, arXiv:1801.03107 

\bibitem[Starkenburg et al.(2016)]{stark16} Starkenburg, T.~K., Helmi, A., \& Sales, L.~V.\ 2016, \aap, 595, A56 

\bibitem[Stewart et al.(2000)]{stewart00} Stewart, S.~G., Fanelli, M.~N., Byrd, G.~G., et al.\ 2000, \apj, 529, 201 

\bibitem[Stierwalt et al.(2015)]{stier15} Stierwalt, S., Besla, G., Patton, D., et al.\ 2015, \apj, 805, 2 

\bibitem[Stierwalt et al.(2017)]{stier17} Stierwalt, S., Liss, S.~E., Johnson, K.~E., et al.\ 2017, arXiv:1701.01731 

\bibitem[Stinson et al.(2007)]{stinson07} Stinson, G.~S., Dalcanton, J.~J., Quinn, T., Kaufmann, T., \& Wadsley, J.\ 2007, \apj, 667, 170 

\bibitem[Tang et al.(2014)]{tang14} Tang, J., Bressan, A., Rosenfield,
  P., et al.\ 2014, \mnras, 445, 4287


\bibitem[Tang et al.(2016)]{tang16} Tang, J., Bressan, A., Slemer, A., et al.\ 2016, \mnras, 455, 3393 

\bibitem[Theis(1999)]{theis99} Theis, C.\ 1999, Reviews in Modern Astronomy, 12, 309 


\bibitem[Tognelli et al.(2011)]{tognelli11} Tognelli, E., Prada Moroni, P.~G., \& Degl'Innocenti, S.\ 2011, \aap, 533, A109 

\bibitem[Toloba et al.(2016)]{toloba16} Toloba, E., Guhathakurta, P., Romanowsky, A.~J., et al.\ 2016, \apj, 824, 35 

\bibitem[Tolstoy et al.(2009)]{THT09} Tolstoy, E., Hill, V., \& Tosi, M.\ 2009, \araa, 47, 371 


\bibitem[Tosi et al.(2001)]{tosi01} Tosi, M., Sabbi, E., Bellazzini, M., et al.\ 2001, \aj, 122, 1271 

\bibitem[Tully(1988)]{tully88} Tully, R.~B.\ 1988, \aj, 96, 73 

\bibitem[van Woerden et al.(1975)]{vanw75} van Woerden, H., Bosma, A., \& Mebold, U.\ 1975, La Dynamique des galaxies spirales, 241, 483 

\bibitem[Weisz et al.(2008)]{weisz08} Weisz, D.~R., Skillman, E.~D., Cannon, J.~M., et al.\ 2008, \apj, 689, 160-183 

\bibitem[Weisz et al.(2009)]{weisz09} Weisz, D.~R., Skillman, E.~D., Cannon, J.~M., et al.\ 2009, \apj, 704, 1538-1569 

\bibitem[Weisz et al.(2011)]{weisz11} Weisz, D.~R., Dalcanton, J.~J., Williams, B.~F., et al.\ 2011, \apj, 739, 5 








\end{thebibliography}

\end{document}